\documentclass[12pt,draftcls,peerreview,onecolumn]{IEEEtran}

\usepackage[dvips]{color}
\usepackage{graphicx}
\usepackage{amsmath}
\begin{document}
%
\title{Exploiting Network Cooperation in Green Wireless Communication}
%
%
%

\author{Yulong~Zou, Jia~Zhu, and Rui~Zhang

\thanks{Y. Zou and J. Zhu are with the Institute of Signal Processing and Transmission, Nanjing University of Posts and Telecommunications, Nanjing, Jiangsu 210003, China.}
\thanks{R. Zhang is with the Electrical and Computer Engineering Department, National University of Singapore, Singapore.}

}

\maketitle

\begin{abstract}
There is a growing interest in energy efficient or so-called ``green'' wireless communication to reduce the energy consumption in cellular networks. Since today's wireless terminals are typically equipped with multiple network access interfaces such as Bluetooth, Wi-Fi, and cellular networks, this paper investigates user terminals cooperating with each other in transmitting their data packets to the base station (BS), by exploiting the multiple network access interfaces, called \emph{inter-network cooperation}. We also examine the conventional schemes without user cooperation and with intra-network cooperation for comparison. Given target outage probability and data rate requirements, we analyze the energy consumption of conventional schemes as compared to the proposed inter-network cooperation by taking into account both physical-layer channel impairments (including path loss, fading, and thermal noise) and upper-layer protocol overheads. It is shown that distances between different network entities (i.e., user terminals and BS) have a significant influence on the energy efficiency of proposed inter-network cooperation scheme. Specifically, when the cooperating users are close to BS or the users are far away from each other, the inter-network cooperation may consume more energy than conventional schemes without user cooperation or with intra-network cooperation. However, as the cooperating users move away from BS and the inter-user distance is not too large, the inter-network cooperation significantly reduces the energy consumption over conventional schemes.

\end{abstract}


\IEEEpeerreviewmaketitle

\section{Introduction}
%
%
%
%
\IEEEPARstart In wireless communication, path loss and fading are two major issues to be addressed in order to improve the quality of service (QoS) of various applications (voice, data, multimedia, etc.) [1]. Typically, the channel path loss and fading are determined by many factors including the terrain environment (urban or rural), electromagnetic wave frequency, distance between the transmitter and receiver, antenna height, and so on. In the case of large path loss and deep fading, more transmit power is generally required to maintain a target QoS requirement. For example, in cellular networks, a user terminal at the edge of its associated cell significantly drains its battery energy much faster than that located at the cell center. Therefore, it is practically important to study energy-efficient or so-called green wireless communication to reduce the energy consumption in cellular networks, especially for cell-edge users [2], [3].

It is worth noting that user cooperation has been recognized as an effective means to achieve spatial diversity. In [4], the authors studied cooperative users in relaying each other's transmission to a common destination and examined the outage performance of various relaying protocols (i.e., fixed relaying, selective relaying, and incremental relaying). In [5], the Alamouti space-time coding was examined for regenerative relay networks, where the relay first decodes its received signals from a source node and then re-encodes and forwards its decoded signal to a destination node. In [6], the authors studied the space-time coding in amplify-and-forward relay networks and proposed a distributed linear dispersion code for the cooperative relay transmissions. More recently, user cooperation has been exploited in emerging cognitive radio networks with various cooperative relaying protocols for spectrum sensing and cognitive transmissions [7], [8]. Generally speaking, a user cooperation protocol consists of two phases: 1) a user terminal broadcasts its signal to its destination and cooperative partners; and 2) the partner nodes forward their received signals to the destination that finally decodes the source message by combining the received multiple signal copies. It is known that user cooperation generally costs orthogonal channel resources to achieve the diversity gain. Moreover, the above-discussed intra-network user cooperation typically operates with a single network access interface.

In today's wireless networks, a user terminal (e.g., smart phone) is typically equipped with multiple network access interfaces to support both short-range communication (via e.g. Bluetooth and Wi-Fi) and long-range communication (via e.g. cellular networks) [9], with different radio characteristics in terms of coverage area and energy consumption. Specifically, the short-range networks provide local-area coverage with low energy consumption, whereas cellular networks offer wider coverage with higher energy consumption. This implies that different radio access networks complement each other in terms of the network coverage and energy consumption. In order to take advantages of different existing radio access networks, it is of high practical interest to exploit the multiple network access interfaces assisted user cooperation, termed inter-network cooperation. In this paper, we study the inter-network cooperation to improve energy efficiency of the cellular uplink transmission with the assistance of a short-range communication network.

The main contributions of this paper are summarized as follows. First, we present an inter-network cooperation framework in a heterogeneous environment consisting of different radio access networks (e.g., a short-range communication network and a cellular network). Then, we examine distributed space-time coding for the proposed scheme and derive its closed-form outage probability in a Rayleigh fading environment. Given target outage probability and data rate requirements, we further pursue an energy consumption analysis by considering both physical-layer channel impairments including path loss, fading, and thermal noise and upper-layer protocol overheads per data packet. For the purpose of comparison, we also examine the energy consumption of two benchmark schemes, including the traditional scheme without user cooperation and existing scheme with intra-network cooperation (i.e., the user terminals cooperate via a common cellular network interface) [5], and show potential advantages of the proposed inter-network cooperation in terms of energy saving.

The remainder of this paper is organized as follows. Section II presents the system model and proposes the inter-network cooperation scheme. In Section III, we present the energy consumption analysis and numerical results of traditional schemes without user cooperation and with intra-network cooperation and proposed inter-network cooperation under uniform outage probability and data rate requirements, which is then extended to a more general scenario with non-uniform outage and rate requirements in Section IV. Finally, Section V provides some concluding remarks and directions for future work.

\section{Cellular Uplink Transmission Based on Network Cooperation}
In this section, we first present the system model of a heterogeneous network environment, where user terminals are assumed to have multiple radio access interfaces including a short-range communication interface and a cellular access interface. Then, we propose the inter-network cooperation scheme by exploiting the short-range network to assist cellular uplink transmissions as well as two baseline schemes for comparison.
\subsection{System Model}
\begin{figure*}
  \centering
  {\includegraphics[scale=0.7]{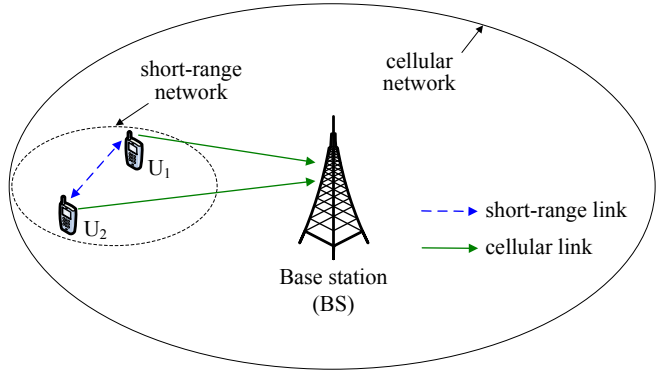}\\
  \caption{A simplified cellular network consisting one base station (BS) and two user terminals that are equipped with multiple radio access interfaces (i.e., a short-range communication interface and a cellular access interface).}\label{Fig1}}
\end{figure*}

Fig. 1 shows the system model of a heterogeneous network consisting of one base station (BS) and two user terminals as denoted by U1 and U2, each equipped with a short-range communication interface and a cellular access interface. The two users are assumed to cooperate with each other in transmitting to BS. Since U1 and U2 are equipped with a short-range communication interface (e.g., Bluetooth), they are able to establish a short-range cooperative network to assist their cellular uplink transmissions and improve the overall energy efficiency. To be specific, we first allow U1 and U2 to communicate with each other and exchange their uplink data packets through the short-range network. Once U1 and U2 obtain each other's data packets, they can employ distributed space-time coding to transmit their data packets to BS by sharing their antennas. Note that the proposed scheme uses two different networks (i.e., a short-range network and a cellular network), which is thus termed inter-network cooperation and differs from traditional user cooperation schemes [4]-[6] that operate in a homogeneous network environment with one single radio access interface. Under the cellular network setup, the traditional user cooperation requires a user terminal to transmit its signal over a cellular frequency band to its partner that then forwards the received signal to BS. This comes at the cost of low cellular spectrum utilization efficiency, since two orthogonal channels are required to complete one packet transmission from a user terminal to BS via its partner. In contrast, the inter-network cooperation allows a user terminal to transmit its signal to its partner using a short-range network (e.g., Bluetooth) over an industrial, scientific and medical (ISM) band, instead of using a cellular band. This thus saves cellular spectrum resources by utilizing the available ISM band and can significantly improve the system performance as compared with the traditional intra-network cooperation. Notice that a more general scenario with multiple user terminals (e.g., more than two users) can be reduced to the two-user cooperation by designing an additional user pairwise grouping protocol [10]. Moreover, different user pairs can proceed with the proposed inter-network cooperation process identically and independently of each other. In addition, similar inter-network cooperation can be applied to cellular downlink transmissions from BS to user terminals. To be specific, BS broadcasts its downlink packets to U1 and U2 over cellular frequency bands. Then, U1 and U2 exchange their received packets through a short-range network so that both U1 and U2 can achieve the cooperative diversity gain without loss of cellular spectrum utilization.
\begin{figure*}
  \centering
  {\includegraphics[scale=0.7]{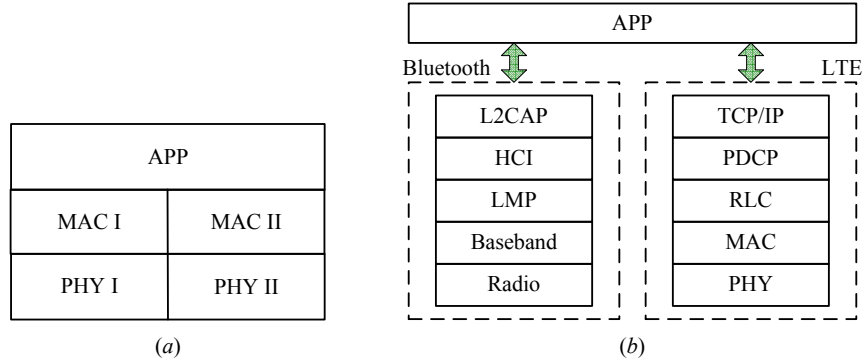}\\
  \caption{{Protocol reference models of the proposed inter-network cooperation: (\emph{a}) a generic model, (\emph{b}) a concrete model considering Bluetooth and LTE.}}\label{Fig2}}
\end{figure*}

Fig. 2 illustrates protocol reference models of the proposed inter-network cooperation, in which a generic model and a more concrete model are shown in Figs. 2(\emph{a}) and 2(\emph{b}), respectively. In Fig. 2(\emph{a}), we assume the use of two sets of MAC and PHY protocols (i.e., MAC I-PHY I and MAC II-PHY II) to implement the inter-network cooperation. Without loss of generality, consider MAC I and PHY I for short-range communication interface and the other MAC-PHY pair (i.e., MAC II and PHY II) for cellular access interface. Therefore, U1 and U2 first exchange their packets through MAC I and PHY I. Then, U1 and U2 assist each other in transmitting their packets to BS by using MAC II and PHY II. One can also see from Fig. 2(a) that the two sets of MAC and PHY share a common application (APP) protocol, implying that the two different network interfaces (i.e., cellular network and short-range network) can be coordinated through the APP protocol. {Furthermore, Fig. 2(\emph{b}) shows a more specific reference model for the proposed inter-network cooperation by considering Bluetooth and long term evolution (LTE) as the short-range and cellular communication interfaces, respectively. As shown in Fig. 2(\emph{b}), the Bluetooth protocol stack consists of the following layers: radio, baseband, link management protocol (LMP), host controller interface (HCI), and logical link control and adaptation protocol (L2CAP) [11]. In contrast, the LTE protocol stack is composed of the PHY, MAC, radio link control (RLC), packet data convergence protocol (PDCP), and TCP/IP protocols [13]. Notice that the interested readers may refer to [11]-[13] for more information about the Bluetooth and LTE protocols.}

\begin{figure*}
  \centering
  {\includegraphics[scale=0.7]{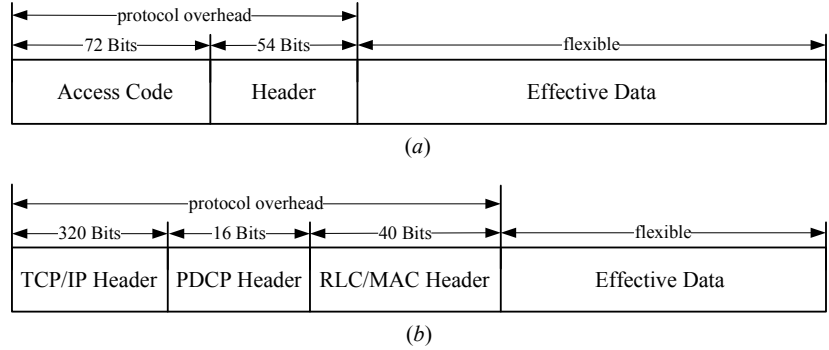}\\
  \caption{{Data packet structures: (\emph{a}) Bluetooth packet frame format, and (\emph{b}) LTE packet frame format.}}\label{Fig3}}
\end{figure*}
{It needs to be pointed out that the data transmission over a radio access network requires certain network overhead including the protocol headers and application-specific information. Moreover, different radio access networks have different protocol architectures and thus have different overhead costs. In Fig. 3, we show the frame structures of Bluetooth and LTE data packets. As shown in Fig. 3(\emph{a}), a standard Bluetooth packet consists of three fields: access code, header, and effective data [12]. In contrast, as shown in Fig. 3(\emph{b}), the LTE packet includes the TCP/IP header, PDCP header, RLC/MAC header, and effective data [13]. Notice that the effective data sizes in both Bluetooth and LTE packets are flexible and vary from zero to thousands of bits. It is worth mentioning that the upper-layer protocol overhead consumes additional energy, which should be taken into consideration for computing the total energy consumption. Therefore, the data rate at physical layer, referred to as \emph{PHY rate} throughout this paper, should consider both the effective data and protocol overhead. In the inter-network cooperation scheme, the cellular and short-range communication interfaces have the same effective data rate. However, as shown in Fig. 3, the protocol overhead in LTE packet differs from that in Bluetooth packet, resulting in the different PHY rates for cellular and short-range communications. Given an effective data rate $\overline R$, the required PHY rate is given by
\begin{equation}\label{equa1}
R=\dfrac{\overline R}{\kappa},
\end{equation}
where $\kappa<1$ is defined as the ratio of the effective data size to the whole packet size, called \emph{effective data ratio}. Considering Bluetooth as the short-range communication interface, the effective data ratio can be obtained from Fig. 3(\emph{a}) as
\begin{equation}\label{equa2}
\kappa_s=\dfrac{N}{N+72+54}=\dfrac{N}{N+126},
\end{equation}
where $N$ is the number of effective bits (Ebits) per date packet denoted by ${\textrm{Ebits/packet}}$ for notational convenience. Meanwhile, from Fig. 3(\emph{b}), we can obtain the effective data ratio of cellular access interface as
\begin{equation}\label{equa3}
\kappa_c=\dfrac{N}{N+320+16+40}=\dfrac{N}{N+376}.
\end{equation}
By integrating network overhead into PHY rate as formulated in (1)-(3), we are able to examine the impact of upper-layer protocol overhead on the energy consumption. }In addition, we consider a general channel model [1] that incorporates the radio frequency, path loss and fading effects in characterizing wireless transmissions, i.e.,
\begin{equation}\label{equa4}
{P_{Rx}} = {P_{Tx}}{\left(\dfrac{\lambda }{{4\pi d}}\right)^2}{G_{Tx}}{G_{Rx}}|h{|^2},
\end{equation}
where ${P_{Rx}}$ is the received power, ${P_{Tx}}$ is the transmitted power, $\lambda $ is the carrier wavelength, $d$ is the transmission distance, ${G_{Tx}}$ is the transmit antenna gain, ${G_{Rx}}$ is the receive antenna gain, and $h$ is the channel fading coefficient. Throughout this paper, we consider a Rayleigh fading model to characterize the channel fading, i.e., $|h{|^2}$ is modeled as an exponential random variable with mean $\sigma _h^2$. Also, all receivers are assumed with the circularly symmetric complex Gaussian (CSCG) distributed thermal noise with zero mean and noise variance $\sigma _n^2$. As is known in [1], the noise variance $\sigma _n^2$ is modeled as $\sigma _n^2 = N_0B$, where $N_0$ is called noise power spectral density in ${\textrm{\emph{dBm/Hz}}}$ and $B$ is the channel bandwidth in $\textrm{\emph{Hz}}$.

\subsection{The Case Without User Cooperation}

\begin{figure*}
  \centering
  {\includegraphics[scale=0.8]{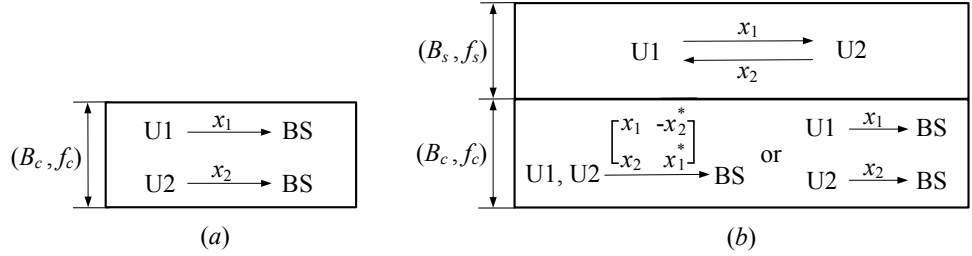}\\
  \caption{{Data transmission diagrams of the traditional scheme without user cooperation and proposed inter-network cooperation for uplink transmissions from U1 and U2 to BS: (\emph{a}) traditional scheme without user cooperation,  and (\emph{b}) proposed inter-network cooperation, where ${f_c}$ and ${B_c}$, respectively, represent the cellular carrier frequency and spectrum bandwidth, and ${f_s}$ and ${B_s}$ are the carrier frequency and spectrum bandwidth of the short-range communications, respectively.}}\label{Fig4}}
\end{figure*}

First, consider the traditional scheme without user cooperation as a baseline for comparison. Without loss of generality, let ${x_1}$ and ${x_2}$ denote transmit signals of U1 and U2, respectively. {Fig. 4 (\emph{a}) shows the traditional cellular uplink transmission process without user cooperation, where ${f_c}$ and ${B_c}$ represent the cellular carrier frequency and spectrum bandwidth, respectively. As shown in Fig. 4(\emph{a}), U1 and U2 transmit their signals ${x_1}$ and ${x_2}$ to BS, respectively, over the cellular spectrum. In cellular networks, BS is regarded as a centralized controller and its associated mobile users access the cellular spectrum with orthogonal multiple access such as time division multiple access (TDMA) and orthogonal frequency division multiple access (OFDMA). Notice that different orthogonal multiple access techniques achieve the same capacity performance in an information-theoretical sense. Throughout this paper, we assume the equal resource allocation between U1 and U2 in accessing cellular spectrum.} Considering that U1 transmits ${x_1}$ with power ${P_1}$ and effective rate ${\overline R_1}$, we can obtain the received signal-to-noise ratio (SNR) at BS from U1 as
\begin{equation}\label{equa5}
\gamma _{1b}^T = \dfrac{{{P_1}}}{{{N_0}{B_c}}}{\left(\dfrac{{{\lambda _c}}}{{4\pi {d_{1b}}}}\right)^2}{G_{U1}}{G_{BS}}|{h_{1b}}{|^2},
\end{equation}
where the superscript $T$ stands for `traditional', ${\lambda _c}=c/f_c$ is the cellular carrier wavelength, $c$ is the speed of light, ${B_c}$ is the cellular spectrum bandwidth, ${d_{1b}}$ is the transmission distance from U1 to BS, ${G_{U1}}$ is the transmit antenna gain at U1, ${G_{BS}}$ is the receive antenna gain at BS, ${h_{1b}}$ is the fading coefficient of the channel from U1 to BS. Similarly, considering that U2 transmits ${x_2}$ with power ${P_2}$ and effective rate ${\overline R_2}$, the received SNR at BS from U2 is given by
\begin{equation}\label{equa6}
\gamma _{2b}^T = \dfrac{{{P_2}}}{{{N_0}{B_c}}}{\left(\dfrac{{{\lambda _c}}}{{4\pi {d_{2b}}}}\right)^2}{G_{U2}}{G_{BS}}|{h_{2b}}{|^2},
\end{equation}
where ${d_{2b}}$ is the transmission distance from U2 to BS, ${G_{U2}}$ is the transmit antenna gain at U2, and ${h_{2b}}$ is the fading coefficient of the channel from U2 to BS.

\subsection{Proposed Inter-Network Cooperation}
In this subsection, we propose the inter-network cooperation scheme for cellular uplink transmissions from U1 and U2 to BS. {Fig. 4(\emph{b}) shows the transmission process of proposed inter-network cooperation, where ${f_s}$ and ${B_s}$, respectively, represent the carrier frequency and spectrum bandwidth of the short-range communications. Differing from the traditional scheme without user cooperation, the inter-network cooperation exploits the short-range communications (between U1 and U2 over frequency $f_s$) to assist the cellular transmissions (from U1 and U2 to BS over frequency $f_c$). The following details the inter-network cooperation process in transmitting ${x_1}$ and ${x_2}$ from U1 and U2 to BS. First, we allow U1 and U2 to exchange their signals over the short-range communication network. The short-range communication is a form of the peer-to-peer communications and two duplex approaches, i.e., time-division duplex (TDD) and frequency-division duplex (FDD), are available to achieve the full-duplex communications between U1 and U2. It is pointed out that both TDD and FDD methods can achieve the same capacity limit in an information-theoretic sense. Similarly to cellular communications, we assume the equal time/frequency allocation between U1 and U2 in accessing the short-range communication spectrum.} Thus, considering that U1 transmits ${x_1}$ to U2 with power ${P_{1,s}}$ and effective rate ${\overline R_1}$ over the short-range communication network, we can obtain the received SNR at U2 from U1 as
\begin{equation}\label{equa7}
\gamma _{12}^{NC} = \dfrac{{{P_{1,s}}}}{{{N_0}{B_s}}}{\left(\dfrac{{{\lambda _s}}}{{4\pi {d_{12}}}}\right)^2}{G_{U1}}{G_{U2}}|{h_{12}}{|^2},
\end{equation}
where the superscript $NC$ stands for `network cooperation', ${\lambda _s}=c/f_s$ is the carrier wavelength of the short-range communication, ${d_{12}}$ is the transmission distance from U1 to U2, and ${h_{12}}$ is the fading coefficient of the channel from U1 to U2. Meanwhile, U2 transmits ${x_2}$ to U1 with power ${P_{2,s}}$ and effective rate ${\overline R_2}$ through the short-range communications, and thus the received SNR at U1 from U2 is written as
\begin{equation}\label{equa8}
\gamma _{21}^{NC} = \dfrac{{{P_{2,s}}}}{{{N_0}{B_s}}}{\left(\dfrac{{{\lambda _s}}}{{4\pi {d_{21}}}}\right)^2}{G_{U1}}{G_{U2}}|{h_{21}}{|^2},
\end{equation}
where ${d_{21}}$ is the transmission distance from U2 to U1, and ${h_{21}}$ is the fading coefficient of the channel from U2 to U1. For notational convenience, let $\theta  = 1$ denote the case that both U1 and U2 succeed in decoding each other's signals through the short-range communications and $\theta  = 2$ denote the other case that either U1 or U2 (or both) fails to decode. In the case of $\theta  = 1$, we adopt Alamouti's space-time code [14], [15] for U1 and U2 in transmitting ${x_1}$ and ${x_2}$ to BS over a cellular band, where the transmit power values of U1 and U2 are denoted by ${P_{1,c}}$ and ${P_{2,c}}$, respectively. The reasons for choosing Alamouti's space-time code are twofold: 1) it is an open loop transmit diversity scheme that does not require channel state information at transmitter; and 2) it is the only space-time code that can achieve the full diversity gain without the loss of data rate. Specifically, given that $\theta=1$ occurs, U1 and U2 transmit ${x_1}$ and ${x_2}$ simultaneously and the received signal at BS is given by
\begin{equation}\label{equa9}
{y_1} = \sqrt {{P_{1,c}}{(\dfrac{{{\lambda _c}}}{{4\pi {d_{1b}}}})^2}{G_{U1}}{G_{BS}}} {h_{1b}}{x_1} + \sqrt {{P_{2,c}}{(\dfrac{{{\lambda _c}}}{{4\pi {d_{2b}}}})^2}{G_{U2}}{G_{BS}}} {h_{2b}}{x_2} + {n_1},
\end{equation}
where ${d_{1b}}$ and ${d_{2b}}$ are the transmission distance from U1 to BS and that from U2 to BS, respectively, ${h_{1b}}$ and ${h_{2b}}$ are fading coefficients of the channel from U1 to BS and that from U2 to BS, respectively, and ${n_{1}}$ is a CSCG random variable with zero mean and noise variance ${N_0}{B_c}$. After that, U1 and U2 transmit $- x_2^*$ and $x_1^*$ simultaneously to BS, where $*$ denotes the conjugate operation. Thus, the signal received at BS is expressed as
\begin{equation}\label{equa10}
{y_2} =  - \sqrt {{P_{1,c}}{(\dfrac{{{\lambda _c}}}{{4\pi {d_{1b}}}})^2}{G_{U1}}{G_{BS}}} {h_{1b}}x_2^* + \sqrt {{P_{2,c}}{(\dfrac{{{\lambda _c}}}{{4\pi {d_{2b}}}})^2}{G_{U2}}{G_{BS}}} {h_{2b}}x_1^* + {n_2},
\end{equation}
where ${n_2}$ is a CSCG random variable with zero mean and noise variance ${N_0}{B_c}$. From (9) and (10), we obtain
\begin{equation}\label{equa11}
\begin{split}
&\left[ {\begin{array}{*{20}{c}}
{\sqrt {{P_{1,c}}{(\dfrac{{{\lambda _c}}}{{4\pi {d_{1b}}}})^2}{G_{U1}}{G_{BS}}} h_{1b}^*}&{\sqrt {{P_{2,c}}{(\dfrac{{{\lambda _c}}}{{4\pi {d_{2b}}}})^2}{G_{U2}}{G_{BS}}} {h_{2b}}}\\
{\sqrt {{P_{2,c}}{(\dfrac{{{\lambda _c}}}{{4\pi {d_{2b}}}})^2}{G_{U2}}{G_{BS}}} h_{2b}^*}&{ - \sqrt {{P_{1,c}}{(\dfrac{{{\lambda _c}}}{{4\pi {d_{1b}}}})^2}{G_{U1}}{G_{BS}}} {h_{1b}}}
\end{array}} \right]\left[ \begin{array}{l}
{y_1}\\
y_2^*
\end{array} \right]\\
 &= [{P_{1,c}}{(\dfrac{{{\lambda _c}}}{{4\pi {d_{1b}}}})^2}{G_{U1}}{G_{BS}}|{h_{1b}}{|^2} + {P_{2,c}}{(\dfrac{{{\lambda _c}}}{{4\pi {d_{2b}}}})^2}{G_{U2}}{G_{BS}}|{h_{2b}}{|^2}]\left[ \begin{array}{l}
{x_1}\\
{x_2}
\end{array} \right]\\
&\quad+ \left[ \begin{array}{l}
\sqrt {{P_{1,c}}{(\dfrac{{{\lambda _c}}}{{4\pi {d_{1b}}}})^2}{G_{U1}}{G_{BS}}} h_{1b}^*{n_1} + \sqrt {{P_{2,c}}{{(\dfrac{{{\lambda _c}}}{{4\pi {d_{2b}}}})}^2}{G_{U2}}{G_{BS}}} {h_{2b}}n_2^*\\
\sqrt {{P_{2,c}}{(\dfrac{{{\lambda _c}}}{{4\pi {d_{2b}}}})^2}{G_{U2}}{G_{BS}}} h_{2b}^*{n_1} - \sqrt {{P_{1,c}}{{(\dfrac{{{\lambda _c}}}{{4\pi {d_{1b}}}})}^2}{G_{U1}}{G_{BS}}} {h_{1b}}n_2^*
\end{array} \right]
\end{split}
\end{equation}
from which BS can decode ${x_1}$ and ${x_2}$ separately. One can observe from (11) that in the case of $\theta  = 1$, the same received SNR is achieved at BS in decoding both ${x_1}$ and ${x_2}$ (from U1 and U2, respectively), which is given by
\begin{equation}\label{equa12}
\begin{split}
\gamma _{1b}^{NC}(\theta  = 1) = \gamma _{2b}^{NC}(\theta  = 1)& = \dfrac{{{P_{1,c}}}}{{{N_0}{B_c}}}{\left(\dfrac{{{\lambda _c}}}{{4\pi {d_{1b}}}}\right)^2}{G_{U1}}{G_{BS}}|{h_{1b}}{|^2}\\
& \quad + \dfrac{{{P_{2,c}}}}{{{N_0}{B_c}}}{\left(\dfrac{{{\lambda _c}}}{{4\pi {d_{2b}}}}\right)^2}{G_{U2}}{G_{BS}}|{h_{2b}}{|^2}.
\end{split}
\end{equation}
In the case of $\theta  = 2$, i.e., either U1 or U2 (or both) fails to decode the short-range transmissions, we allow U1 and U2 to transmit their signals ${x_1}$ and ${x_2}$ to BS separately over a cellular band by using an orthogonal multiple access method (e.g., TDMA or OFDMA). Therefore, in the case of $\theta  = 2$, the received SNRs at BS in decoding ${x_1}$ and ${x_2}$ (from U1 and U2) are, respectively, given by
\begin{equation}\label{equa13}
\gamma _{1b}^{NC}(\theta  = 2) = \dfrac{{{P_{1,c}}}}{{{N_0}{B_c}}}{\left(\dfrac{{{\lambda _c}}}{{4\pi {d_{1b}}}}\right)^2}{G_{U1}}{G_{BS}}|{h_{1b}}{|^2},
\end{equation}
and
\begin{equation}\label{equa14}
\gamma _{2b}^{NC}(\theta  = 2) = \dfrac{{{P_{2,c}}}}{{{N_0}{B_c}}}{\left(\dfrac{{{\lambda _c}}}{{4\pi {d_{2b}}}}\right)^2}{G_{U2}}{G_{BS}}|{h_{2b}}{|^2}.
\end{equation}
This completes the signal model of the inter-network cooperation scheme.

\subsection{Conventional Intra-Network Cooperation}
For the purpose of comparison, this subsection presents the conventional intra-network cooperation scheme [4], [5]. Similarly, we consider U1 and U2 that transmit ${x_1}$ and ${x_2}$ to BS, respectively. In the conventional intra-network cooperation scheme [5], U1 and U2 first exchange their signals (i.e., ${x_1}$ and ${x_2}$) between each other over cellular bands, which is different from the inter-network cooperation case where the information exchanging operates in a short-range communication network over ISM bands. During the information exchange process, U1 and U2 attempt to decode each other's signals. If both U1 and U2 successfully decode, the Alamouti space-time coding is used in transmitting ${x_1}$ and ${x_2}$ from U1 and U2 to BS over cellular bands. Otherwise, U1 and U2 transmit ${x_1}$ and ${x_2}$ to BS, separately. Note that the conventional scheme requires two orthogonal phases to complete each packet transmission, which causes the loss of one-half of cellular spectrum utilization. Thus, the conventional scheme needs to transmit at twice of the data rate of the inter-network cooperation scheme in order to send the same amount of information. In other words, U1 and U2 should transmit ${x_1}$ and ${x_2}$ at effective rates $2{\overline R_1}$ and $2\overline R_2$, respectively, for a fair comparison. One can observe that the signal model of the conventional intra-network cooperation is almost the same as that of the inter-network cooperation, except that $2{\overline R_1}$ and $2\overline R_2$ are considered as the effective rates of U1 and U2 in the conventional scheme and, moreover, the information exchange between U1 and U2 in the conventional scheme operates over cellular bands instead of ISM bands.

\section{Energy Consumption Analysis with Uniform Outage and Rate Requirements}
In this section, we analyze the energy consumption of the traditional scheme without user cooperation, the conventional intra-network cooperation, and the proposed inter-network cooperation assuming that different users (i.e., U1 and U2) have the same outage probability and data rate requirements, called uniform outage and rate requirements. Furthermore, we present numerical results based on the energy consumption analysis to show the advantage of the proposed scheme over the two baseline schemes under certain conditions.
\subsection{The Case Without User Cooperation}
Without loss of generality, let $\overline {{\rm{Pout}}} $ and $\overline R $ denote the common target outage probability and effective data rate, respectively, for both users. {From (1)-(3), we obtain the PHY rates of short-range and cellular communications as $R_s=\overline R/\kappa_s$ and $R_c=\overline R/\kappa_c$. Due to the limited error correction capability in practical communication systems, both the short-range and cellular communications cannot achieve the Shannon capacity [16]. Moreover, the cellular communications typically has more powerful error-correcting capability than the short-range communications. Therefore, let $\Delta_s$ and $\Delta_c$ denote performance gaps for the short-range communications from the capacity limit and for the cellular communications from the capacity limit, respectively.} Using (5) and considering a performance gap $\Delta_c$ away from Shannon capacity [16], we obtain the maximum achievable rate from U1 to BS of the traditional scheme without user cooperation as
\begin{equation}\label{equa15}
C_{1b}^T = {B_c}{\log _2}\left(1 + \dfrac{\gamma _{1b}^T}{\Delta_c }\right) = {B_c}{\log _2}\left[ 1 + \dfrac{{{P_1}}}{{\Delta_c{N_0}{B_c}}}{(\dfrac{{{\lambda _c}}}{{4\pi {d_{1b}}}})^2}{G_{U1}}{G_{BS}}|{h_{1b}}{|^2} \right],
\end{equation}
where $\Delta_c>1$. As we know, an outage event occurs when the channel capacity falls below the data rate. Note that the random variable $|{h_{1b}}{|^2}$ follows an exponential distribution with mean $\sigma _{1b}^2$. Thus, we can compute the outage probability for U1's transmission as
\begin{equation}\label{equa16}
{\rm{Pout}}_1^T = \Pr \left(C_{1b}^T < \dfrac{\overline R}{\kappa_c}\right) = 1 - \exp \left[ - \dfrac{{16{\pi ^2}\Delta_c{N_0}{B_c}d_{1b}^2({2^{{\overline R}/({B_c}\kappa_c)}} - 1)}}{{{P_1}\sigma _{1b}^2{G_{U1}}{G_{BS}}\lambda _c^2}} \right].
\end{equation}
Given the target outage probability $\overline {{\rm{Pout}}} $ (i.e., ${\rm{Pout}}_1^T = \overline {{\rm{Pout}}} $), we can easily compute the power consumption of U1 from (16) as
\begin{equation}\nonumber
{P_1} =  - \dfrac{{16{\pi ^2}\Delta_c{N_0}{B_c}d_{1b}^2({2^{\overline R /({B_c}\kappa_c)}} - 1)}}{{\sigma _{1b}^2{G_{U1}}{G_{BS}}\lambda _c^2\ln (1 - \overline {{\rm{Pout}}} )}}.
\end{equation}
Typically, user terminals are powered by battery and the battery discharging is shown as a nonlinear process [17]. Specifically, given the transmit pulse power $P_1$ and pulse duration $T_p$, the battery energy consumption of U1 is given by
\begin{equation}\label{equa17}
{E_1} = \frac{{{{(1 + \varepsilon )}^2}\xi\omega  }}{{V{\eta ^2}}}{({P_1}{T_p})^2} + \frac{{(1 + \varepsilon )}}{\eta }{P_1}{T_p} + \frac{{{P_c}{T_p}}}{\eta },
\end{equation}
where $\xi\buildrel \Delta \over = \int_0^{{T_p}} {{{[{p_0}(t)]}^2}dt} $ is determined by a so-called normalized transmit pulse shape ${p_0}(t) = p(t)/\int_0^{{T_p}} {|p(t){|^2}dt} $, $\omega$ is the battery efficiency factor, $V$ is the battery voltage, $\eta$ is the transfer efficiency of the DC-to-DC converter, $\varepsilon$ is the extra power loss factor of the power amplifier, and $P_c$ is the circuit power consumption. Assuming that U1 and U2 have the same transmit pulse shape and circuit power consumption, we can similarly obtain the battery energy consumption of U2 as
\begin{equation}\label{equa18}
{E_2} = \frac{{{{(1 + \varepsilon )}^2}\xi\omega  }}{{V{\eta ^2}}}{({P_2}{T_p})^2} + \frac{{(1 + \varepsilon )}}{\eta }{P_2}{T_p} + \frac{{{P_c}{T_p}}}{\eta },
\end{equation}
where $P_2$ is given by
\begin{equation}\nonumber
{P_2} =  - \dfrac{{16{\pi ^2}\Delta_c{N_0}{B_c}d_{2b}^2({2^{\overline R /({B_c}\kappa_c)}} - 1)}}{{\sigma _{2b}^2{G_{U2}}{G_{BS}}\lambda _c^2\ln (1 - \overline {{\rm{Pout}}} )}}.
\end{equation}
Therefore, given the case without user cooperation, the total battery energy consumption of both U1 and U2 is obtained as
\begin{equation}\label{equa19}
{E_T} = {E_1} + {E_2},
\end{equation}
where $E_1$ and $E_2$ are given by (17) and (18), respectively.
\subsection{Proposed Inter-Network Cooperation}
In this subsection, we present an energy consumption analysis for the proposed inter-network cooperation. Notice that all random variables $|{h_{12}}{|^2}$, $|{h_{21}}{|^2}$, $|{h_{1b}}{|^2}$ and $|{h_{2b}}{|^2}$ follow independent exponential distributions with means $\sigma _{12}^2$, $\sigma _{21}^2$, $\sigma _{1b}^2$ and $\sigma _{2b}^2$, respectively. Using (7) and considering a performance gap $\Delta_s$ away from Shannon capacity, we can obtain an outage probability of the short-range transmission from U1 to U2 as
\begin{equation}\label{equa20}
{\rm{Pou}}{{\rm{t}}_{12}} = \Pr \left(\dfrac{\gamma _{12}^{NC}}{\Delta_s} < {2^{{\overline R}/({B_s}\kappa_s)}} - 1\right) = 1 - \exp \left[ - \dfrac{{16{\pi ^2}\Delta_s{N_0}{B_s}d_{12}^2({2^{{\overline R}/({B_s}\kappa_s)}} - 1)}}{{{P_{1,s}}\sigma _{12}^2{G_{U1}}{G_{U2}}\lambda _s^2}} \right].
\end{equation}
Assuming ${\rm{Pou}}{{\rm{t}}_{12}} = \overline {{\rm{Pout}}} $, we can obtain $P_{1,s}$ from the preceding equation as
\begin{equation}\label{equa21}
{P_{1,s}} =  - \dfrac{{16{\pi ^2}\Delta_s{N_0}{B_s}d_{12}^2({2^{\overline R /({B_s}\kappa_s)}} - 1)}}{{\sigma _{12}^2{G_{U1}}{G_{U2}}\lambda _s^2\ln (1 - \overline {{\rm{Pout}}} )}},
\end{equation}
from which the battery energy consumption of U1 for short-range communication is given by
\begin{equation}\label{equa22}
{E_{1,s}} = \frac{{{{(1 + \varepsilon )}^2}\xi\omega  }}{{V{\eta ^2}}}{({P_{1,s}}{T_p})^2} + \frac{{(1 + \varepsilon )}}{\eta }{P_{1,s}}{T_p} + \frac{{{P_c}{T_p}}}{\eta }.
\end{equation}
From (8), we similarly obtain the battery energy consumption of U2 for short-range communication as
\begin{equation}\label{equa23}
{E_{2,s}} = \frac{{{{(1 + \varepsilon )}^2}\xi\omega  }}{{V{\eta ^2}}}{({P_{2,s}}{T_p})^2} + \frac{{(1 + \varepsilon )}}{\eta }{P_{2,s}}{T_p} + \frac{{{P_c}{T_p}}}{\eta },
\end{equation}
where ${P_{2,s}}$ is given by
\begin{equation}\label{equa24}
{P_{2,s}} =  - \dfrac{{16{\pi ^2}\Delta_s{N_0}{B_s}d_{21}^2({2^{\overline R /({B_s}\kappa_s)}} - 1)}}{{\sigma _{21}^2{G_{U1}}{G_{U2}}\lambda _s^2\ln (1 - \overline {{\rm{Pout}}} )}}.
\end{equation}
In addition, considering the target PHY rate $R_c=\overline R/\kappa_c$ for cellular communications, we obtain the outage probability of U1's transmission with the inter-network cooperation from (12) and (13) as
\begin{equation}\label{equa25}
\begin{split}
{\rm{Pout}}_1^{NC} =& \Pr (\theta  = 1)\Pr [\gamma _{1b}^{NC}(\theta  = 1) < ({2^{{\overline R}/({B_c}\kappa_c)}} - 1)\Delta_c] \\
&+ \Pr (\theta  = 2)\Pr [\gamma _{1b}^{NC}(\theta  = 2) < ({2^{{\overline R}/({B_c}\kappa_c)}} - 1)\Delta_c].
\end{split}
\end{equation}
As discussed in Section II-B, case $\theta  = 1$ implies that both U1 and U2 succeed in decoding each other's signals through short-range communications, and $\theta  = 2$ means that either U1 or U2 (or both) fails to decode in the short-range transmissions. Considering the target PHY rate $R_s=\overline R/\kappa_s$ for short-range communications, we can describe $\theta  = 1$ and $\theta  = 2$ as follows
\begin{equation}\label{equa26}
\begin{split}
&\theta  = 1:{\textrm{  }}{B_s}{\log _2}\left(1 + \dfrac{\gamma _{12}^{NC}}{\Delta_s}\right) > \dfrac{{\overline R}}{\kappa_s}{\textrm{  and  }}{B_s}{\log _2}\left(1 + \dfrac{\gamma _{21}^{NC}}{\Delta_s}\right) > \dfrac{{\overline R}}{\kappa_s}\\
&\theta  = 2:{\textrm{  }}{B_s}{\log _2}\left(1 + \dfrac{\gamma _{12}^{NC}}{\Delta_s}\right) < \dfrac{{\overline R}}{\kappa_s}{\textrm{    or   }}{B_s}{\log _2}\left(1 + \dfrac{\gamma _{21}^{NC}}{\Delta_s}\right) < \dfrac{{\overline R}}{\kappa_s}.
\end{split}
\end{equation}
Assuming $\overline {{\rm{Pout}}} $ for short-range communication between U1 and U2, we have
\begin{equation}\label{equa27}
\Pr (\theta  = 1) = \Pr \left(\dfrac{\gamma _{12}^{NC}}{\Delta_s} > {2^{{\overline R}/({B_s}\kappa_s)}} - 1\right)\Pr \left(\dfrac{\gamma _{21}^{NC}}{\Delta_s}  > {2^{{\overline R}/({B_s}\kappa_s)}} - 1\right) = {(1 - \overline {{\rm{Pout}}} )^2},
\end{equation}
and
\begin{equation}\label{equa28}
\Pr (\theta  = 2) = 1 - {(1 - \overline {{\rm{Pout}}} )^2}.
\end{equation}
Since random variables $|{h_{1b}}{|^2}$ and $|{h_{2b}}{|^2}$ are independent and both follow exponential distributions with respective mean $\sigma _{1b}^2$ and $\sigma _{2b}^2$, we can substitute (12) into $\Pr [\gamma _{1b}^{NC}(\theta  = 1) < ({2^{{\overline R}/({B_c}\kappa_c)}} - 1)\Delta_c]$ and obtain
\begin{equation}\label{equa29}
\begin{split}
&\Pr [\gamma _{1b}^{NC}(\theta  = 1) < ({2^{{\overline R}/({B_c}\kappa_c)}} - 1)\Delta_c]\\
&= \Pr [\frac{{{P_{1,c}}}}{{\Delta_c{N_0}{B_c}}}{(\frac{{{\lambda _c}}}{{4\pi {d_{1b}}}})^2}{G_{U1}}{G_{BS}}|{h_{1b}}{|^2} + \frac{{{P_{2,c}}}}{{\Delta_c{N_0}{B_c}}}{(\frac{{{\lambda _c}}}{{4\pi {d_{2b}}}})^2}{G_{U2}}{G_{BS}}|{h_{2b}}{|^2} < {2^{{\overline R}/({B_c}\kappa_c)}} - 1]\\
&= \left\{ \begin{split}
&\begin{split}
&1 - [1 + \frac{{16{\pi ^2}\Delta_c{N_0}{B_c}d_{1b}^2({2^{{\overline R}/({B_c}\kappa_c)}} - 1)}}{{\sigma _{1b}^2{P_{1,c}}\lambda _c^2{G_{U1}}{G_{BS}}}}]\\
&\quad\quad\times\exp \left[ - \frac{{16{\pi ^2}\Delta_c{N_0}{B_c}d_{1b}^2({2^{{\overline R}/({B_c}\kappa_c)}} - 1)}}{{\sigma _{1b}^2{P_{1,c}}\lambda _c^2{G_{U1}}{G_{BS}}}} \right]
\end{split},\quad\quad\sigma _{1b}^2{P_{1,c}}d_{1b}^{ - 2}{G_{U1}} = \sigma _{2b}^2{P_{2,c}}d_{2b}^{ - 2}{G_{U2}}{\rm{    }}\\
&\begin{split}
&1 - \frac{{\sigma _{1b}^2{P_{1,c}}d_{1b}^{ - 2}{G_{U1}}}}{{\sigma _{1b}^2{P_{1,c}}d_{1b}^{ - 2}{G_{U1}} - \sigma _{2b}^2{P_{2,c}}d_{2b}^{ - 2}{G_{U2}}}}\exp \left[ - \frac{{16{\pi ^2}\Delta_c{N_0}{B_c}d_{1b}^2({2^{{\overline R}/({B_c}\kappa_c)}} - 1)}}{{\sigma _{1b}^2{P_{1,c}}\lambda _c^2{G_{U1}}{G_{BS}}}} \right]\\
&{\textrm{  }} - \frac{{\sigma _{2b}^2{P_{2,c}}d_{2b}^{ - 2}{G_{U2}}}}{{\sigma _{2b}^2{P_{2,c}}d_{2b}^{ - 2}{G_{U2}} - \sigma _{1b}^2{P_{1,c}}d_{1b}^{ - 2}{G_{U1}}}}\exp \left[ - \frac{{16{\pi ^2}\Delta_c{N_0}{B_c}d_{2b}^2({2^{{\overline R}/({B_c}\kappa_c)}} - 1)}}{{\sigma _{2b}^2{P_{2,c}}\lambda _c^2{G_{U2}}{G_{BS}}}} \right]
\end{split},{\textrm{    otherwise}}
\end{split} \right.
\end{split}.
\end{equation}
Besides, using (13), we can easily obtain $\Pr [\gamma _{1b}^{NC}(\theta  = 2) <( {2^{{\overline R}/({B_c}\kappa_c)}} - 1)\Delta_c]$ as
\begin{equation}\label{equa30}
\Pr [\gamma _{1b}^{NC}(\theta  = 2) < ({2^{{\overline R}/({B_c}\kappa_c)}} - 1)\Delta_c] = 1 - \exp \left[ - \dfrac{{16{\pi ^2}\Delta_c{N_0}{B_c}d_{1b}^2({2^{{\overline R}/({B_c}\kappa_c)}} - 1)}}{{{P_{1,c}}\sigma _{1b}^2{G_{U1}}{G_{BS}}\lambda _c^2}} \right].
\end{equation}
Similarly to (25), the outage probability of U2's transmissions is obtained from (12) and (14) as
\begin{equation}\label{equa31}
\begin{split}
{\rm{Pout}}_2^{NC} = &\Pr (\theta  = 1)\Pr [\gamma _{2b}^{NC}(\theta  = 1) < ({2^{{\overline R}/({B_c}\kappa_c)}} - 1)\Delta_c] \\
&+ \Pr (\theta  = 2)\Pr [\gamma _{2b}^{NC}(\theta  = 2) < ({2^{{\overline R}/({B_c}\kappa_c)}} - 1)\Delta_c],
\end{split}
\end{equation}
where $\Pr (\theta  = 1)$ and $\Pr (\theta  = 2)$ are given by (27) and (28), respectively. Moreover, the corresponding $\Pr [\gamma _{2b}^{NC}(\theta  = 1) < ({2^{{\overline R}/({B_c}\kappa_c)}} - 1)\Delta_c]$ and $\Pr [\gamma _{2b}^{NC}(\theta  = 2) < ({2^{{\overline R}/({B_c}\kappa_c)}} - 1)\Delta_c]$ are given by
\begin{equation}\label{equa32}
\begin{split}
&\Pr [\gamma _{2b}^{NC}(\theta  = 1) < ({2^{{\overline R}/({B_c}\kappa_c)}} - 1)\Delta_c]\\
&= \left\{ \begin{split}
&\begin{split}
&1 - [1 + \frac{{16{\pi ^2}\Delta_c{N_0}{B_c}d_{1b}^2({2^{{\overline R}/({B_c}\kappa_c)}} - 1)}}{{\sigma _{1b}^2{P_{1,c}}\lambda _c^2{G_{U1}}{G_{BS}}}}]\\
&\quad\quad\times\exp \left[ - \frac{{16{\pi ^2}\Delta_c{N_0}{B_c}d_{1b}^2({2^{{\overline R}/({B_c}\kappa_c)}} - 1)}}{{\sigma _{1b}^2{P_{1,c}}\lambda _c^2{G_{U1}}{G_{BS}}}} \right]
\end{split},\quad\quad\sigma _{1b}^2{P_{1,c}}d_{1b}^{ - 2}{G_{U1}} = \sigma _{2b}^2{P_{2,c}}d_{2b}^{ - 2}{G_{U2}}{\rm{ }}\\
&\begin{split}
&1 - \frac{{\sigma _{1b}^2{P_{1,c}}d_{1b}^{ - 2}{G_{U1}}}}{{\sigma _{1b}^2{P_{1,c}}d_{1b}^{ - 2}{G_{U1}} - \sigma _{2b}^2{P_{2,c}}d_{2b}^{ - 2}{G_{U2}}}}\exp \left[ - \frac{{16{\pi ^2}\Delta_c{N_0}{B_c}d_{1b}^2({2^{{\overline R}/({B_c}\kappa_c)}} - 1)}}{{\sigma _{1b}^2{P_{1,c}}\lambda _c^2{G_{U1}}{G_{BS}}}} \right]\\
&{\textrm{  }}- \frac{{\sigma _{2b}^2{P_{2,c}}d_{2b}^{ - 2}{G_{U2}}}}{{\sigma _{2b}^2{P_{2,c}}d_{2b}^{ - 2}{G_{U2}} - \sigma _{1b}^2{P_{1,c}}d_{1b}^{ - 2}{G_{U1}}}}\exp \left[ - \frac{{16{\pi ^2}\Delta_c{N_0}{B_c}d_{2b}^2({2^{{\overline R}/({B_c}\kappa_c)}} - 1)}}{{\sigma _{2b}^2{P_{2,c}}\lambda _c^2{G_{U2}}{G_{BS}}}} \right]
 \end{split},{\textrm{    otherwise}}
\end{split} \right.
\end{split},
\end{equation}
and
\begin{equation}\label{equa33}
\Pr [\gamma _{2b}^{NC}(\theta  = 2) < ({2^{{\overline R}/({B_c}\kappa_c)}} - 1)\Delta_c] = 1 - \exp \left[ - \frac{{16{\pi ^2}\Delta_c{N_0}{B_c}d_{2b}^2({2^{{\overline R}/({B_c}\kappa_c)}} - 1)}}{{{P_{2,c}}\sigma _{2b}^2{G_{U2}}{G_{BS}}\lambda _c^2}} \right].
\end{equation}
Considering ${\rm{Pout}}_1^{NC} = {\rm{Pout}}_2^{NC} = \overline {{\rm{Pout}}} $ and using (25) and (31), we have
\begin{equation}\label{equa34}
\Pr [\gamma _{1b}^{NC}(\theta  = 2) < ({2^{\overline R /({B_c}\kappa_c)}} - 1)\Delta_c] = \Pr [\gamma _{2b}^{NC}(\theta  = 2) < ({2^{\overline R /({B_c}\kappa_c)}} - 1)\Delta_c],
\end{equation}
which can be further simplified to
\begin{equation}\label{equa35}
{P_{2,c}} = \frac{{\sigma _{1b}^2{G_{U1}}d_{2b}^2}}{{\sigma _{2b}^2{G_{U2}}d_{1b}^2}}{P_{1,c}}.
\end{equation}
Therefore, letting ${\rm{Pout}}_1^{NC} = \overline {{\rm{Pout}}} $ and substituting (35) into (32), we obtain the following equation from (25) as
\begin{equation}\label{equa36}
\begin{split}
&[1 + \frac{{16{\pi ^2}\Delta_c{N_0}{B_c}d_{1b}^2({2^{\overline R /({B_c}\kappa_c)}} - 1){{(1 - \overline {{\rm{Pout}}} )}^2}}}{{\sigma _{1b}^2{P_{1,c}}\lambda _c^2{G_{U1}}{G_{BS}}}}]\\
&\times\exp \left[ - \frac{{16{\pi ^2}\Delta_c{N_0}{B_c}d_{1b}^2({2^{\overline R /({B_c}\kappa_c)}} - 1)}}{{\sigma _{1b}^2{P_{1,c}}\lambda _c^2{G_{U1}}{G_{BS}}}} \right]= 1 - \overline {{\rm{Pout}}} .
\end{split}
\end{equation}
By denoting $x =  - \frac{1}{{{{(1 - \overline {{\rm{Pout}}} )}^2}}} - \frac{{16{\pi ^2}\Delta_c{N_0}{B_c}d_{1b}^2({2^{\overline R /({B_c}\kappa_c)}} - 1)}}{{\sigma _{1b}^2{P_{1,c}}\lambda _c^2{G_{U1}}{G_{BS}}}}$, the preceding equation is written as
\begin{equation}\nonumber
x\exp (x) = \frac{{\exp [ - {{(1 - \overline {{\rm{Pout}}} )}^{ - 2}}]}}{{\overline {{\rm{Pout}}}  - 1}},
\end{equation}
from which $x$ can be solved as
\begin{equation}\label{equa37}
x = W\left( {\frac{{\exp [ - {{(1 - \overline {{\rm{Pout}}} )}^{ - 2}}]}}{{\overline {{\rm{Pout}}}  - 1}}} \right),
\end{equation}
where $W( \cdot )$ is the lambert function. Substituting $x =  - \frac{1}{{{{(1 - \overline {{\rm{Pout}}} )}^2}}} - \frac{{16{\pi ^2}\Delta_c{N_0}{B_c}d_{1b}^2({2^{\overline R /({B_c}\kappa_c)}} - 1)}}{{\sigma _{1b}^2{P_{1,c}}\lambda _c^2{G_{U1}}{G_{BS}}}}$ into (37) yields
\begin{equation}\label{equa38}
{P_{1,c}} = \frac{{16{\pi ^2}\Delta_c{N_0}{B_c}d_{1b}^2({2^{\overline R /({B_c}\kappa_c)}} - 1)}}{{\sigma _{1b}^2\lambda _c^2{G_{U1}}{G_{BS}}}}{\left[ { - \frac{1}{{{{(1 - \overline {{\rm{Pout}}} )}^2}}} - W\left( {\frac{{\exp [ - {{(1 - \overline {{\rm{Pout}}} )}^{ - 2}}]}}{{\overline {{\rm{Pout}}}  - 1}}} \right)} \right]^{ - 1}},
\end{equation}
from which the battery energy consumption of U1 for cellular transmissions is easily given by
\begin{equation}\label{equa39}
{E_{1,c}} = \frac{{{{(1 + \varepsilon )}^2}\xi\omega  }}{{V{\eta ^2}}}{({P_{1,c}}{T_p})^2} + \frac{{(1 + \varepsilon )}}{\eta }{P_{1,c}}{T_p} + \frac{{{P_c}{T_p}}}{\eta }.
\end{equation}
Meanwhile, combining (35) and (38), we have
\begin{equation}\label{equa40}
{P_{2,c}} = \frac{{16{\pi ^2}\Delta_c{N_0}{B_c}d_{2b}^2({2^{\overline R /({B_c}\kappa_c)}} - 1)}}{{\sigma _{2b}^2\lambda _c^2{G_{U2}}{G_{BS}}}}{\left[ { - \frac{1}{{{{(1 - \overline {{\rm{Pout}}} )}^2}}} - W\left( {\frac{{\exp [ - {{(1 - \overline {{\rm{Pout}}} )}^{ - 2}}]}}{{\overline {{\rm{Pout}}}  - 1}}} \right)} \right]^{ - 1}},
\end{equation}
which leads to the battery energy consumption of U2 for cellular transmissions as given by
\begin{equation}\label{equa41}
{E_{2,c}} = \frac{{{{(1 + \varepsilon )}^2}\xi\omega  }}{{V{\eta ^2}}}{({P_{2,c}}{T_p})^2} + \frac{{(1 + \varepsilon )}}{\eta }{P_{2,c}}{T_p} + \frac{{{P_c}{T_p}}}{\eta }.
\end{equation}
Notice that in case of $\theta  = 1$, the Alamouti space-time coding is employed, resulting in that a total battery energy of $2({E_{1,c}} + {E_{2,c}})$ is consumed by U1 and U2 in transmitting to BS. In case of $\theta  = 2$, U1 and U2 consume a total battery energy of $({E_{1,c}} + {E_{2,c}})$ in transmitting to BS. Therefore, considering both the short-range communication and cellular transmissions, the total energy consumption by the inter-network cooperation is given by
\begin{equation}\label{equa42}
\begin{split}
{E_{NC}} &= {E_{1,s}} + {E_{2,s}} + 2\Pr (\theta  = 1)({E_{1,c}} + {E_{2,c}}) + \Pr (\theta  = 2)({E_{1,c}} + {E_{2,c}})\\
&= {E_{1,s}} + {E_{2,s}} + [1+(1-\overline {{\rm{Pout}}})^2]({E_{1,c}} + {E_{2,c}}), \\
\end{split}
\end{equation}
where ${E_{1,s}}$ and ${E_{2,s}}$ are given by (22) and (23), respectively, while ${E_{1,c}}$ and ${E_{2,c}}$ are given by (39) and (41), respectively.
\subsection{Conventional Intra-Network Cooperation}
As discussed in Section II-D, the intra-network cooperation differs from the proposed inter-network cooperation in two main aspects. First, in the intra-network cooperation case, U1 and U2 should transmit at effective rate $2\overline R$ to send the same amount of information as the inter-network cooperation case. This means that in characterizing the energy consumption of intra-network cooperation, we need to replace $\overline R$ in the energy consumption expressions of the inter-network cooperation, i.e., (21), (24), (38) and (40), with $2\overline R$. Secondly, in the intra-network cooperation, the information exchange between U1 and U2 operates over cellular bands instead of ISM bands. Therefore, the energy consumed in the information exchange phase of intra-network cooperation can be obtained from (21) and (24) by replacing $\lambda_s$ and $B_s$ with $\lambda_c$ and $B_c$, respectively. In this way, the total energy consumption of conventional intra-network cooperation can be readily determined.

\subsection{Numerical Results}
\begin{table*}
  \centering
  \caption{{System Parameters Used in Numerical Examples.}}
  {\includegraphics[scale=0.9]{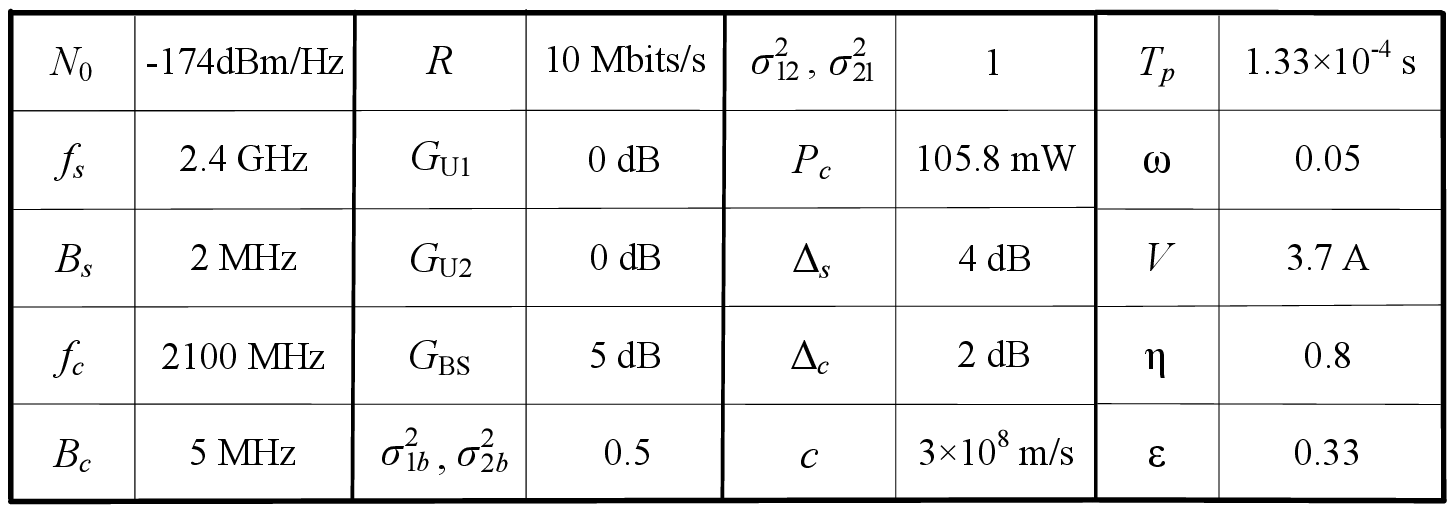}\label{Tab1}}
\end{table*}

In this subsection, we present numerical results on the battery energy consumption over a pulse interval $T_p$ of various transmission schemes given the uniform target outage probability and effective rate requirements, i.e., $\overline {{\rm{Pout}}}  = {10^{ - 4}}$ and $\overline R  = 10M{\textrm{bits/s}}$. Table I summarizes the system parameters used in the numerical evaluations, where ${f_s} = 2.4{\textrm{GHz}}$ and ${B_s} = 2{\textrm{MHz}}$ correspond to a Bluetooth low energy (BLE) system that operates at $2.4 {\textrm{GHz}}$ with $40$ channels of $2 {\textrm{MHz}}$ each. The cellular carrier frequency ${f_c} = 2100{\textrm{MHz}}$ and cellular bandwidth ${B_s} = 5{\textrm{MHz}}$ are typically considered in 3GPP long term evolution (LTE) currently operating at $2100 {\textrm{MHz}}$ in North America with various options of channel bandwidth including $1.4 {\textrm{MHz}}$, $3 {\textrm{MHz}}$, $5 {\textrm{MHz}}$, $10 {\textrm{MHz}}$, $15 {\textrm{MHz}}$ and $20 {\textrm{MHz}}$. The antenna gains of U1 and U2 are set as ${G_{U1}} = {G_{U2}} = 0{\textrm{dB}}$ and the BS's antenna gain ${G_{BS}} = 5{\textrm{dB}}$ is assumed. {Considering the fact that cellular communications typically has more powerful error-correcting capability than short-range communications, the performance gaps $\Delta_s$ and $\Delta_c$ of short-range and cellular communications from their respective Shannon limits are given by $\Delta_s=4{\textrm{dB}}$ and $\Delta_c=2{\textrm{dB}}$.} In addition, the remaining parameters in Table I are specified according to [17].
\begin{figure*}
  \centering
  {\includegraphics[scale=0.85]{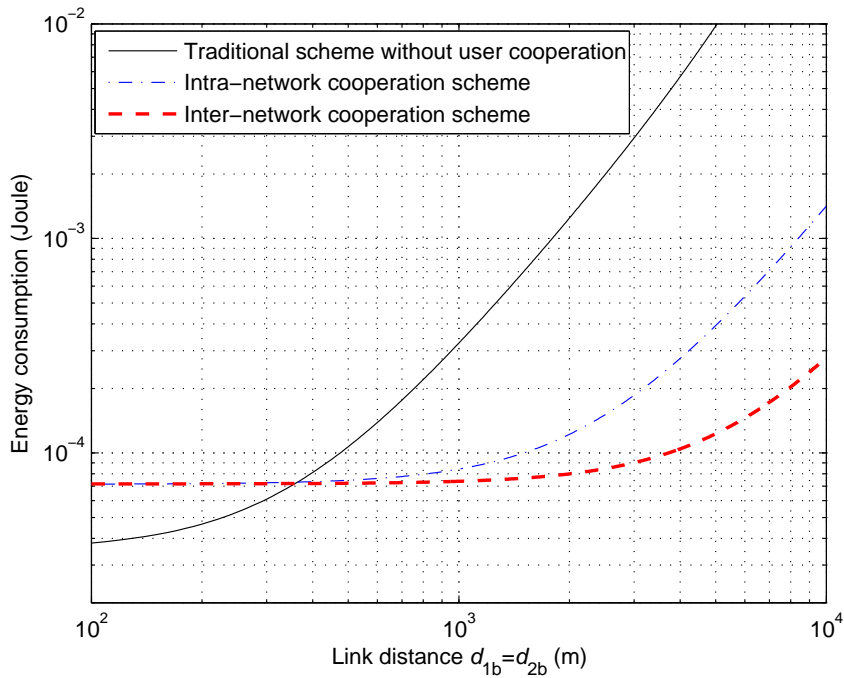}\\
  \caption{{Energy consumption versus link distance from U1/U2 to BS, ${d_{1b}}$ and ${d_{2b}}$ in meters ($m$), of various transmission schemes with target outage probability $\overline {{\rm{Pout}}}  = {10^{ - 4}}$, effective rate $\overline R  = 10M{\textrm{bits/s}}$, $N=2000\textrm{Ebits/packet}$, and inter-user distance ${d_{12}} = {d_{21}} = 20m$.}}\label{Fig5}}
\end{figure*}

Fig. 5 shows the energy consumption comparison among the traditional scheme without user cooperation, the intra-network cooperation, and the inter-network cooperation by plotting (19) and (42) with target outage probability $\overline {{\rm{Pout}}}  = {10^{ - 4}}$, effective rate $\overline R  = 10M{\textrm{bits/s}}$, $N=2000\textrm{Ebits/packet}$, and ${d_{12}} = {d_{21}} = 20m$. As shown in Fig. 5, with short link distances (e.g., $d_{1b},d_{2b} \l 300m$), both the intra- and inter-network cooperation consume more energy than the traditional scheme without user cooperation. This implies that it is not beneficial to exploit user cooperation when user terminals are close enough to BS. However, as the link distances from U1 and U2 to BS (i.e., ${d_{1b}}$ and ${d_{2b}}$) both increase beyond certain value, the energy consumptions of the intra- and inter-network cooperation become lower than that of the scheme without user cooperation. Fig. 5 shows that, when $d_{1b}$ and $d_{2b}$ are both larger than $400m$, the intra- and inter-network cooperation outperform the non-cooperative counterpart in terms of energy consumption. This means that the user cooperation can significantly save energy when users move away from BS (e.g., at the cell edge). In addition, it is observed from Fig. 5 that the inter-network cooperation always outperforms the intra-network cooperation in terms of energy saving, showing the advantage of inter-network cooperation over intra-network cooperation.
\begin{figure*}
  \centering
  {\includegraphics[scale=0.85]{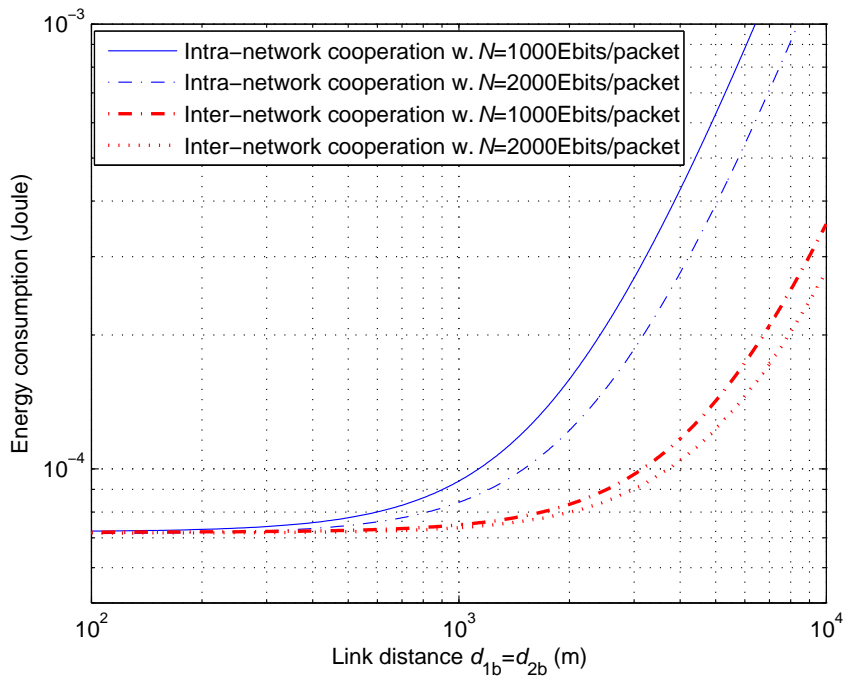}\\
  \caption{{Energy consumption versus link distance from U1/U2 to BS, ${d_{1b}}$ and ${d_{2b}}$ in meters ($m$), of various transmission schemes for different number of effective bits (Ebits) per data packet $N$ with target outage probability $\overline {{\rm{Pout}}}  = {10^{ - 4}}$, effective rate $\overline R  = 10M{\textrm{bits/s}}$, and inter-user distance ${d_{12}} = {d_{21}} = 20m$.}}\label{Fig6}}
\end{figure*}

{Fig. 6 illustrates the energy consumption versus link distances from U1 and U2 to BS of the conventional intra-network cooperation and proposed inter-network cooperation for different number of effective bits (Ebits) per packet (i.e., $N=1000{\textrm{Ebits/packet}}$ and $N=2000{\textrm{Ebits/packet}}$). As shown in Fig. 6, the inter-network cooperation scheme always performs better than the intra-network cooperation in terms of energy consumption for both cases of $N=1000{\textrm{Ebits/packet}}$ and $N=2000{\textrm{Ebits/packet}}$. However, this energy saving becomes less notable, as ${d_{1b}}$ and ${d_{2b}}$ both decrease. The reason is that when ${d_{1b}}$ and ${d_{2b}}$ decrease, ${P_{1,c}}$ and ${P_{2,c}}$ decrease and the information exchange between U1 and U2 in turn accounts for an increasing part of the total energy consumption. Moreover, the information exchange in intra-network cooperation over cellular bands consumes less energy than that of inter-network cooperation wherein the short-range communication is used for information exchange over ISM bands, which results in that the energy saving of the inter-network cooperation becomes less significant as ${d_{1b}}$ and ${d_{2b}}$ decrease. In addition, as the number of Ebits per packet $N$ increases from $N=1000{\textrm{Ebits/packet}}$ to $N=2000{\textrm{Ebits/packet}}$, the energy consumptions of both intra- and inter-network cooperation decrease. This is due to the fact that increasing the number of Ebits per packet reduces the upper-layer protocol overhead percentage in a data packet and thus saves the energy of transmitting protocol overhead, leading to the energy consumption reduction with an increasing number of Ebits per packet $N$.}
\begin{figure*}
  \centering
  {\includegraphics[scale=0.85]{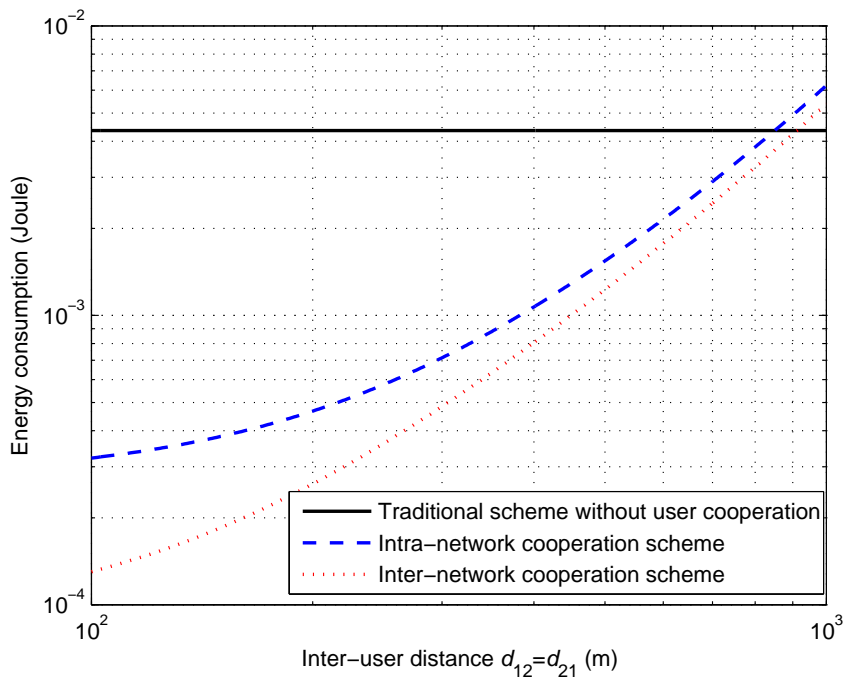}\\
  \caption{{Energy consumption versus inter-user distance between U1 and U2 of various transmission schemes with target outage probability $\overline {{\rm{Pout}}}  = {10^{ - 4}}$, effective rate $\overline R  = 10M{\textrm{bits/s}}$, $N=1000\textrm{Ebits/packet}$, ${d_{1b}}=3000m$, and ${d_{2b}}=3100m$.}}\label{Fig7}}
\end{figure*}

Fig. 7 shows the energy consumption versus inter-user distance between U1 and U2 of the traditional scheme without user cooperation, the conventional intra-network cooperation, and the proposed inter-network cooperation with target outage probability $\overline {{\rm{Pout}}}  = {10^{ - 4}}$, effective rate $\overline R  = 10M{\textrm{bits/s}}$, $N=1000\textrm{Ebits/packet}$, ${d_{1b}}=3000m$, and ${d_{2b}}=3100m$. As shown in Fig. 7, the energy consumption of traditional scheme without user cooperation is constant in this case, which is due to the fact that ${E_T}$ given by (19) is independent of inter-user distance. It can be observed from Fig. 7 that when the inter-user distance is relatively small, both the intra- and inter-network cooperation significantly outperform the traditional scheme without user cooperation in terms of energy consumption. However, as the inter-user distance increases beyond a certain value, the intra- and inter-network cooperation perform worse than the traditional scheme without user cooperation, showing that user cooperation is not energy saving when U1 and U2 are far away from each other. It is also shown from Fig. 7 that the inter-network cooperation strictly outperforms the intra-network cooperation in terms of energy consumption.
\begin{figure*}
  \centering
  {\includegraphics[scale=0.85]{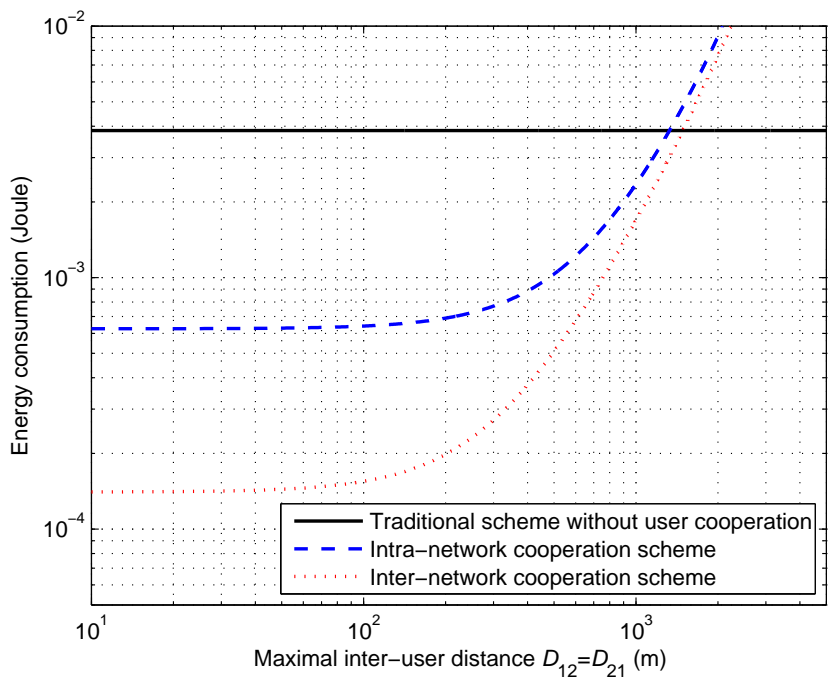}\\
  \caption{{Average energy consumption versus ${D_{12}}={D_{21}}$ of various transmission schemes with target outage probability $\overline {{\rm{Pout}}}  = {10^{ - 4}}$, effective rate $\overline R  = 10M{\textrm{bits/s}}$, $N=1000\textrm{Ebits/packet}$, and ${D_{1b}}={D_{2b}}=5000m$.}}\label{Fig8}}
\end{figure*}

Considering the fact that users may move around in cellular networks and link distances vary over time, we now investigate the impact of random link distances on the battery energy consumption. As observed from (21), (24), (38) and (40), the consumed powers $P_{1,s}$, $P_{2,s}$, $P_{1,c}$ and $P_{2,c}$ are proportional to link distances $d^2_{12}$, $d^2_{21}$, $d^2_{1b}$, and $d^2_{2b}$, respectively. Without loss of generality, we assume that the link distances $d_{12}$, $d_{21}$, $d_{1b}$ and $d_{2b}$ follow independent and uniform distributions, i.e., ${d_{12}} \sim U\left( {0,{D_{12}}} \right)$, ${d_{21}} \sim U\left( {0,{D_{21}}} \right)$, ${d_{1b}} \sim U\left( {0,{D_{1b}}} \right)$, and ${d_{2b}} \sim U\left( {0,{D_{2b}}} \right)$. One can easily obtain the expected distances $\bar d_{12}^2$, $\bar d_{21}^2$, $\bar d_{1b}^2$ and $\bar d_{2b}^2$ as $D_{12}^2/3$, $D_{21}^2/3$, $D_{1b}^2/3$, and $D_{2b}^2/3$, respectively. Hence, given the random distances ${d_{12}} \sim U\left( {0,{D_{12}}} \right)$, ${d_{21}} \sim U\left( {0,{D_{21}}} \right)$, ${d_{1b}} \sim U\left( {0,{D_{1b}}} \right)$, and ${d_{2b}} \sim U\left( {0,{D_{2b}}} \right)$, the average power consumption $\bar P_{1,s}$, $\bar P_{2,s}$, $\bar P_{1,c}$ and $\bar P_{2,c}$ can be obtained by substituting $d_{12}^2=D_{12}^2/3$, $d_{21}^2=D_{21}^2/3$, $d_{1b}^2=D_{1b}^2/3$, and $d_{2b}^2=D_{2b}^2/3$ into (21), (24), (38) and (40), respectively. Fig. 8 shows the average energy consumption of the traditional scheme without user cooperation, the conventional intra-network cooperation, and the proposed inter-network cooperation with target outage probability $\overline {{\rm{Pout}}}  = {10^{ - 4}}$, effective rate $\overline R  = 10M{\textrm{bits/s}}$, $N=1000\textrm{Ebits/packet}$, and ${D_{1b}}={D_{2b}}=5000m$. It is shown from Fig. 8 that even considering random link distances, the proposed inter-network cooperation is more energy efficient than both the traditional scheme without user cooperation and the intra-network cooperation as long as $D_{12}$ and $D_{21}$ are relatively small (e.g., $D_{12}, D_{21} < 10^3$).

\section{Energy Consumption Analysis with Non-Uniform Outage and Rate Requirements}
In this section, we extend the energy consumption analysis of the proposed inter-network cooperation to the more general case that U1 and U2 have different outage probability and data rate requirements, termed non-uniform outage and rate requirements. Without loss of generality, let ${\overline {{\rm{Pout}}} _i}$ and ${\overline R _i}$, respectively, denote the target outage probability and effective data rate of the transmission from ${\textrm{U}}_i$ to BS, $i = 1,2$. {From (1)-(3), we obtain the PHY rates of U1 and U2 with the short-range communications as $R_{1,s}=\overline R_1/\kappa_s$ and $R_{2,s}=\overline R_2/\kappa_s$. Similarly, the PHY rates of U1 and U2 with the cellular communications are given by $R_{1,c}=\overline R_1/\kappa_c$ and $R_{2,c}=\overline R_2/\kappa_c$.} Besides, the target outage probability of the transmission from U1 to U2 and that from U2 to U1 are in general not identical, denoted by ${\rm{Pou}}{{\rm{t}}_{12}} = {\overline {{\rm{Pout}}} _{12}}$ and ${\rm{Pou}}{{\rm{t}}_{21}} = {\overline {{\rm{Pout}}} _{21}}$, respectively. Thus, considering ${\rm{Pou}}{{\rm{t}}_{12}} = {\overline {{\rm{Pout}}} _{12}}$ and ${R_{1,s}} = {\overline R _1}/\kappa_s$ and using (7), we can obtain the power consumption of U1 for the short-range communication as
\begin{equation}\nonumber
{P_{1,s}} =  - \frac{{16{\pi ^2}\Delta_s{N_0}{B_s}d_{12}^2({2^{{{\overline R }_1}/({B_s}\kappa_s)}} - 1)}}{{\sigma _{12}^2{G_{U1}}{G_{U2}}\lambda _s^2\ln (1 - {{\overline {{\rm{Pout}}} }_{12}})}},
\end{equation}
from which the battery energy consumption of U1 for the short-range communications is given by
\begin{equation}\label{equa43}
{E_{1,s}} = \frac{{{{(1 + \varepsilon )}^2}\xi\omega  }}{{V{\eta ^2}}}{({P_{1,s}}{T_p})^2} + \frac{{(1 + \varepsilon )}}{\eta }{P_{1,s}}{T_p} + \frac{{{P_c}{T_p}}}{\eta }.
\end{equation}
Similarly, the power consumption of U2 for the short-range communication with target outage probability ${\rm{Pou}}{{\rm{t}}_{21}} = {\overline {{\rm{Pout}}} _{21}}$ and PHY rate ${R_{2,s}} = {\overline R _2/\kappa_s}$ is given by
\begin{equation}\nonumber
{P_{2,s}} =  - \frac{{16{\pi ^2}\Delta_s{N_0}{B_s}d_{21}^2({2^{{{\overline R }_2}/({B_s}\kappa_s)}} - 1)}}{{\sigma _{21}^2{G_{U1}}{G_{U2}}\lambda _s^2\ln (1 - {{\overline {{\rm{Pout}}} }_{21}})}},
\end{equation}
which results in the battery energy consumption of U2 for the short-range communications as given by
\begin{equation}\label{equa44}
{E_{2,s}} = \frac{{{{(1 + \varepsilon )}^2}\xi\omega  }}{{V{\eta ^2}}}{({P_{2,s}}{T_p})^2} + \frac{{(1 + \varepsilon )}}{\eta }{P_{2,s}}{T_p} + \frac{{{P_c}{T_p}}}{\eta }.
\end{equation}
In addition, considering ${\rm{Pou}}{{\rm{t}}_{12}} = {\overline {{\rm{Pout}}} _{12}}$, ${\rm{Pou}}{{\rm{t}}_{21}} = {\overline {{\rm{Pout}}} _{21}}$, ${R_{1,c}} = {\overline R _1}/\kappa_c$ and ${R_{2,c}} = {\overline R _2}/\kappa_c$, we obtain the outage probability of U1's transmissions with the inter-network cooperation from (25) as
\begin{equation}\label{equa45}
\begin{split}
{\rm{Pout}}_1^{NC} =& (1 - {\overline {{\rm{Pout}}} _{12}})(1 - {\overline {{\rm{Pout}}} _{21}})\Pr [\gamma _{1b}^{NC}(\theta  = 1) < ({2^{{{\overline R }_1}/({B_c}\kappa_c)}} - 1)\Delta_c]\\
& + ({\overline {{\rm{Pout}}} _{12}} + {\overline {{\rm{Pout}}} _{21}} - {\overline {{\rm{Pout}}} _{12}}{\overline {{\rm{Pout}}} _{21}})\Pr [\gamma _{1b}^{NC}(\theta  = 2) < ({2^{{{\overline R }_1}/({B_c}\kappa_c)}} - 1)\Delta_c].
\end{split}
\end{equation}
Similarly, the outage probability of U2's transmissions is obtained from (31) as
\begin{equation}\label{equa46}
\begin{split}
{\rm{Pout}}_2^{NC} =& (1 - {\overline {{\rm{Pout}}} _{12}})(1 - {\overline {{\rm{Pout}}} _{21}})\Pr [\gamma _{2b}^{NC}(\theta  = 1) < ({2^{{{\overline R }_2}/({B_c}\kappa_c)}} - 1)\Delta_c]\\
& + ({\overline {{\rm{Pout}}} _{12}} + {\overline {{\rm{Pout}}} _{21}} - {\overline {{\rm{Pout}}} _{12}}{\overline {{\rm{Pout}}} _{21}})\Pr [\gamma _{2b}^{NC}(\theta  = 2) < ({2^{{{\overline R }_2}/({B_c}\kappa_c)}} - 1)\Delta_c].
\end{split}
\end{equation}
Therefore, the transmit power of U1 and U2 for the cellular transmissions (i.e., ${P_{1,c}}$ and ${P_{2,c}}$) should be minimized subject to the target outage and rate requirements, which can be modeled as the following optimization problem
\begin{equation}\label{equa47}
\begin{split}
&\mathop {\min }\limits_{{P_{1,c}},{P_{2,c}}} {P_{1,c}} + {P_{2,c}}\\
&\begin{split}
{\textrm{s}}{\textrm{.t}}{\textrm{.    }}&\alpha \Pr [\gamma _{1b}^{NC}(\theta  = 1) < ({2^{{{\overline R }_1}/({B_c}\kappa_c)}} - 1)\Delta_c]\\
 &\quad + (1 - \alpha )\Pr [\gamma _{1b}^{NC}(\theta  = 2) < ({2^{{{\overline R }_1}/({B_c}\kappa_c)}} - 1)\Delta_c] \le {\overline {{\rm{Pout}}} _1}\\
&\alpha \Pr [\gamma _{2b}^{NC}(\theta  = 1) < ({2^{{{\overline R }_2}/({B_c}\kappa_c)}} - 1)\Delta_c] \\
&\quad + (1 - \alpha )\Pr [\gamma _{2b}^{NC}(\theta  = 2) < ({2^{{{\overline R }_2}/({B_c}\kappa_c)}} - 1)\Delta_c] \le {\overline {{\rm{Pout}}} _2}\\
&{P_{1,c}} \ge 0,{\rm{  }}{P_{2,c}} \ge 0,
\end{split}
\end{split}
\end{equation}
where $\alpha  = (1 - {\overline {{\rm{Pout}}} _{12}})(1 - {\overline {{\rm{Pout}}} _{21}})$. It is difficult to solve the above problem due to the non-convex constraints involving $\Pr [\gamma _{1b}^{NC}(\theta  = 1) < ({2^{{{\overline R }_1}/({B_c}\kappa_c)}} - 1)\Delta_c]$ and $\Pr [\gamma _{2b}^{NC}(\theta  = 1) < ({2^{{{\overline R }_2}/({B_c}\kappa_c)}} - 1)\Delta_c]$ as shown in (29) and (32). For the purpose of exposition, we examine the solution for this problem by assuming the high SNR values of ${P_{1,c}}/{N_0}{B_c}$ and ${P_{2,c}}/{N_0}{B_c}$ in order to simplify the expressions of $\Pr [\gamma _{1b}^{NC}(\theta  = 1) < ({2^{{{\overline R }_1}/({B_c}\kappa_c)}} - 1)\Delta_c]$ and $\Pr [\gamma _{2b}^{NC}(\theta  = 1) < ({2^{{{\overline R }_2}/({B_c}\kappa_c)}} - 1)\Delta_c]$. Denoting ${\gamma _{1,c}} = \frac{{{P_{1,c}}}}{{{N_0}{B_c}}}$, ${\gamma _{2,c}} = \frac{{{P_{2,c}}}}{{{N_0}{B_c}}}$, ${K_1} = {(\frac{{{\lambda _c}}}{{4\pi {d_{1b}}}})^2}{G_{U1}}{G_{BS}}$, ${K_2} = {(\frac{{{\lambda _c}}}{{4\pi {d_{2b}}}})^2}{G_{U2}}{G_{BS}}$, ${z_1} = \frac{{{\gamma _{1,c}}\sigma _{1b}^2{K_1}}}{{({2^{{{\overline R }_1}/({B_c}\kappa_c)}} - 1)\Delta_c}}$ and ${z_2} = \frac{{{\gamma _{2,c}}\sigma _{2b}^2{K_2}}}{{({2^{{{\overline R }_1}/({B_c}\kappa_c)}} - 1)\Delta_c}}$, and using the two-dimensional Taylor approximation with a given accuracy of $\delta > 0$, we obtain (see Appendix A for the details)
\begin{equation}\label{equa48}
\Pr [\gamma _{1b}^{NC}(\theta  = 1) < ({2^{{{\overline R }_1}/({B_c}\kappa_c)}} - 1)\Delta_c]  \approx  \frac{1}{{2{z_1}{z_2}}},
\end{equation}
where $0 < z_1^{ - 1} + z_2^{ - 1} < \sqrt {2\delta } $. Similarly, $\Pr [\gamma _{2b}^{NC}(\theta  = 1) < ({2^{{{\overline R }_2}/({B_c}\kappa_c)}} - 1)\Delta_c]$ is approximated as
\begin{equation}\label{equa49}
\Pr [\gamma _{2b}^{NC}(\theta  = 1) < ({2^{{{\overline R }_2}/({B_c}\kappa_c)}} - 1)\Delta_c] \approx \frac{1}{{2{\beta ^2}{z_1}{z_2}}},
\end{equation}
where $\beta  = \frac{{{2^{{{\overline R }_1}/({B_c}\kappa_c)}} - 1}}{{{2^{{{\overline R }_2}/({B_c}\kappa_c)}} - 1}}$ and $0 < z_1^{ - 1} + z_2^{ - 1} < \sqrt {2\delta } \beta $. In addition, assuming large values of ${z_1}$ and ${z_2}$ and applying the one-dimensional Taylor approximation to (30) and (33) with an accuracy $\delta $, we obtain
\begin{figure*}
  \centering
  {\includegraphics[scale=0.85]{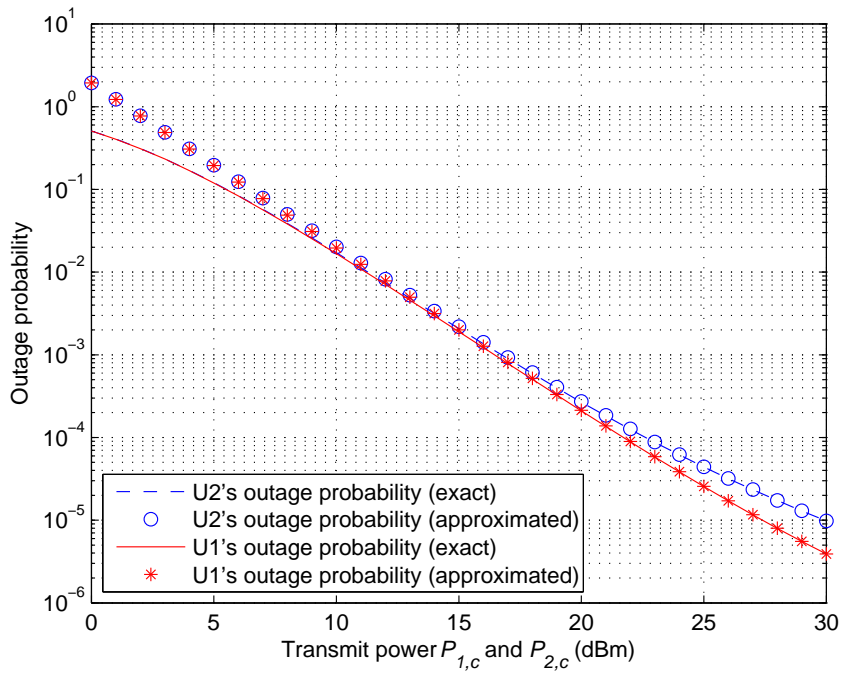}\\
  \caption{{Exact and approximate outage probabilities of U1 and U2 versus transmit powers $P_{1,c}$ and $P_{2,c}$ with $\overline {{\rm{Pout}}}_{12}  = \overline {{\rm{Pout}}}_{21}={10^{ - 3}}$, $\overline R_1 =\overline R_2  = 5M{\textrm{bits/s}}$, $N=1000\textrm{Ebits/packet}$, ${d_{1b}}=2000m$, and ${d_{2b}}=4000m$.}}\label{Fig9}}
\end{figure*}
\begin{equation}\label{equa50}
\begin{split}
&\Pr [\gamma _{1b}^{NC}(\theta  = 2) < ({2^{{R_1}/({B_c}\kappa_c)}} - 1)\Delta_c] \approx \frac{1}{{{z_1}}}\\
&\Pr [\gamma _{2b}^{NC}(\theta  = 2) < ({2^{{R_2}/({B_c}\kappa_c)}} - 1)\Delta_c] \approx \frac{1}{{{z_2}}},
\end{split}
\end{equation}
where $0 < z_1^{ - 1},z_2^{ - 1} < \sqrt {2\delta } $. In order to show the effectiveness of the Taylor approximations as given by (48)-(50), Fig. 9 illustrates exact and approximate outage probabilities of U1 and U2 versus transmit powers $P_{1,c}$ and $P_{2,c}$ with $\overline {{\rm{Pout}}}_{12}  = \overline {{\rm{Pout}}}_{21}={10^{ - 3}}$, $\overline R_1 =\overline R_2  = 5M{\textrm{bits/s}}$, $N=1000\textrm{Ebits/packet}$, ${d_{1b}}=2000m$, and ${d_{2b}}=4000m$. One can observe from Fig. 9 that for both U1 and U2, the exact and approximate outage probabilities match well with each other, especially when the transmit powers $P_{1,c}$ and $P_{2,c}$ are larger than $10{\textrm{dBm}}$. This means that the exact outage probabilities of U1 and U2 can be well approximated by using (48)-(50). Thus, substituting (48)-(50) into (47), we have
\begin{equation}\label{equa51}
\begin{split}
&\mathop {\max }\limits_{{z_1},{z_2}}  - {\rho _1}{z_1} - {\rho _2}{z_2}\\
&\begin{split}
{\textrm{s}}{\textrm{.t}}{\textrm{. }}&2{\overline {{\rm{Pout}}} _1}{z_1}{z_2} - 2(1 - \alpha ){z_2} - \alpha  \ge 0\\
&2{\overline {{\rm{Pout}}} _2}{\beta ^2}{z_1}{z_2} - 2(1 - \alpha ){\beta ^2}{z_1} - \alpha  \ge 0\\
&0 < z_1^{ - 1} + z_2^{ - 1} < \min (\sqrt {2\delta } ,\sqrt {2\delta } \beta )\\
&0 < z_1^{ - 1},z_2^{ - 1} < \sqrt {2\delta },
\end{split}
\end{split}
\end{equation}
where ${\rho _1} = \frac{{\Delta_c{N_0}{B_c}({2^{{{\overline R }_1}/({B_c}\kappa_c)}} - 1)}}{{\sigma _{1b}^2{K_1}}}$ and ${\rho _2} = \frac{{\Delta_c{N_0}{B_c}({2^{{{\overline R }_1}/({B_c}\kappa_c)}} - 1)}}{{\sigma _{2b}^2{K_2}}}$. From Appendix B, we show that (51) is a non-convex problem. The following presents a so-called ``heuristic" approach to determine the optimal solution to any non-convex problem. Although the optimization problem as formulated in (51) is non-convex, its optimal solution still needs to satisfy the Karush-Kuhn-Tucker (KKT) necessary condition [18]. Thus, there exists  $(\mu _1^*,\mu _2^*)$ and $(z_1^*, z_2^*)$ such that the first-order KKT condition is met, i.e.,
\begin{equation}\label{equa52}
\left\{ \begin{array}{l}
2({\overline {{\rm{Pout}}} _1}\mu _1^* + {\overline {{\rm{Pout}}} _2}{\beta ^2}\mu _2^*)z_2^* - 2(1 - \alpha ){\beta ^2}\mu _2^* - {\rho _1} = 0\\
2({\overline {{\rm{Pout}}} _1}\mu _1^* + {\overline {{\rm{Pout}}} _2}{\beta ^2}\mu _2^*)z_1^* - 2(1 - \alpha )\mu _1^* - {\rho _2} = 0\\
\mu _1^*[2{\overline {{\rm{Pout}}} _1}z_1^*z_2^* - 2(1 - \alpha )z_2^* - \alpha ] = 0\\
\mu _2^*[2{\overline {{\rm{Pout}}} _2}{\beta ^2}z_1^*z_2^* - 2(1 - \alpha ){\beta ^2}z_1^* - \alpha ] = 0\\
\mu _1^*,\mu _2^* \ge 0
\end{array} \right..
\end{equation}
Using (52), we may obtain multiple solutions satisfying the first-order necessary condition and then need to verify whether the solutions achieve local optimums by checking the second-order sufficient condition. Now, let us recall the second-order sufficient condition as follows. Denoting $z=(z_1,z_2)^T$ and $\mu=(\mu_1,\mu_2)^T$, consider the problem to maximize $J(z)=- {\rho _1}{z_1} - {\rho _2}{z_2}$ subject to the constraints $f(z)=[f_1(z),f_2(z)]^T$ wherein $f_1(z)=2{\overline {{\rm{Pout}}} _1}{z_1}{z_2} - 2(1 - \alpha ){z_2} - \alpha = 0$ and $f_2(z)=2{\overline {{\rm{Pout}}} _2}{\beta ^2}{z_1}{z_2} - 2(1 - \alpha ){\beta ^2}{z_1} - \alpha = 0$. Letting $L(z,\mu)=J(z)+\mu^Tf(z)$, we have a strict local maximum $z^*=(z_1^*,z_2^*)$, if the following equations hold
\begin{equation}\label{equa53}
\begin{split}
&{\nabla _z}L({z^*},{\mu ^*}) = 0\\
&{\nabla _\mu }L({z^*},{\mu ^*}) = 0\\
&{y^T}\nabla _{zz}^2L({z^*},{\mu ^*})y < 0,\forall y=(y_1,y_2)^T \ne 0{\textrm{ such that }}\nabla f{({z^*})^T}y = 0.
\end{split}
\end{equation}
Therefore, given a solution to (52), we first determine the active inequality constraints, and then validate whether or not the solution is a local maximum by using (53). After that, we can choose the global maximum among the multiple local maximums as the optimal solution (i.e., $z_1^*$ and $z_2^*$). Once $z_1^*$ and $z_2^*$ are obtained, the transmit power of U1 and U2 for the cellular transmissions (i.e., ${P_{1,c}}$ and ${P_{2,c}}$) are given by
\begin{equation}\label{equa54}
{P_{1,c}} = \frac{{\Delta_c{N_0}{B_c}({2^{{{\overline R }_1}/({B_c}\kappa_c)}} - 1)}}{{\sigma _{1b}^2{K_1}}}z_1^*,
\end{equation}
and
\begin{equation}\label{equa55}
{P_{2,c}} = \frac{{\Delta_c{N_0}{B_c}({2^{{{\overline R }_1}/({B_c}\kappa_c)}} - 1)}}{{\sigma _{2b}^2{K_2}}}z_2^*.
\end{equation}
From (54) and (55), we can easily obtain the battery energy consumption of U1 and U2 for the cellular transmissions as
\begin{equation}\label{equa56}
{E_{1,c}} = \frac{{{{(1 + \varepsilon )}^2}\xi\omega  }}{{V{\eta ^2}}}{({P_{1,c}}{T_p})^2} + \frac{{(1 + \varepsilon )}}{\eta }{P_{1,c}}{T_p} + \frac{{{P_c}{T_p}}}{\eta },
\end{equation}
and
\begin{equation}\label{equa57}
{E_{2,c}} = \frac{{{{(1 + \varepsilon )}^2}\xi\omega  }}{{V{\eta ^2}}}{({P_{2,c}}{T_p})^2} + \frac{{(1 + \varepsilon )}}{\eta }{P_{2,c}}{T_p} + \frac{{{P_c}{T_p}}}{\eta }.
\end{equation}
It is pointed out that although the energy consumption given in (56) and (57) is nonlinear in transmit power $P_{1,c}$ and $P_{2,c}$, we here consider the minimization of $P_{1,c}+P_{2,c}$ as the objective function in (47) by ignoring second-order terms of the battery energy consumption. Hence, combining (43), (44), (56), and (57), the total battery energy consumption of the inter-network cooperation scheme is given by
\begin{equation}\label{equa58}
{E_{NC}} = {E_{1,s}} + {E_{2,s}} + (1 + \alpha )({E_{1,c}} + {E_{2,c}}).
\end{equation}

Next, we provide numerical results on the battery energy consumption of the inter-network cooperation scheme with non-uniform outage and rate requirements. Table II shows the total energy consumptions of the traditional scheme without user cooperation and proposed inter-network cooperation, while for brevity, the case of intra-network cooperation is skipped here. Also, the optimal solutions $(z_1^*,z_2^*)$ are obtained by using the proposed heuristic method and the exhaustive search. As shown in the second row of Table II, the optimal solutions $(z_1^*,z_2^*)$ obtained from the heuristic method and the exhaustive search match well with each other. It is also observed from the third and fourth rows of Table II that, as the target outage probabilities decrease, the transmit power of U1 and U2 both increases accordingly, i.e., the more stringent the outage probability requirements are, the more transmit power is required. In addition, one can see that the total energy consumed by the proposed scheme, ${E_{NC}}$, is smaller than that by the traditional scheme without user cooperation, ${E_T}$. This further verifies the advantage of the proposed scheme in terms of energy consumption even under the setup of non-uniform QoS requirements.
\begin{table*}
  \centering
  \caption{{Energy Consumption Comparison between the Traditional Scheme Without User Cooperation and Proposed Inter-Network Cooperation with ${\overline R _1} = 25M{\textrm{bits/s}}$, ${\overline R _2} = 15M{\textrm{bits/s}}$, $N=1000{\textrm{Ebits/packet}}$, ${d_{1b}} = 8400m$, ${d_{2b}} = 8500m$, and ${d_{12}} = {d_{21}} = 150m$.}}
  {\includegraphics[scale=0.85]{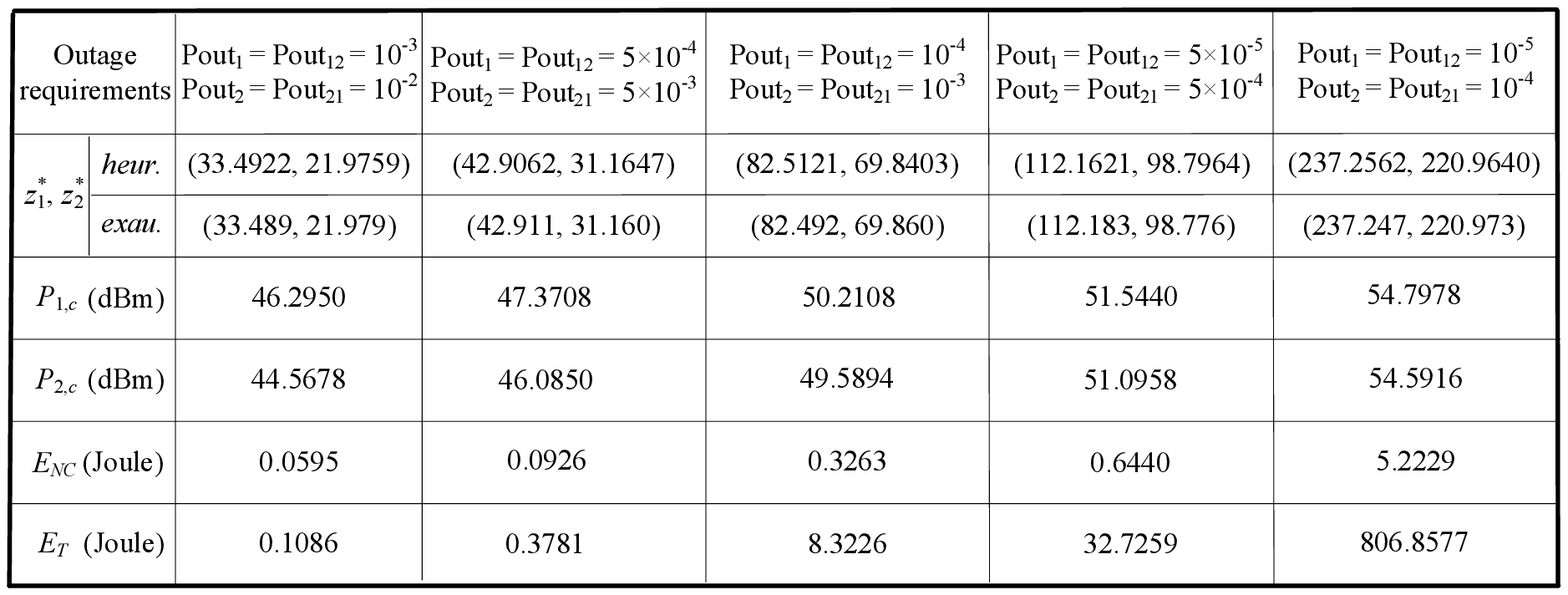}\label{Tab2}}
\end{table*}

\section{Conclusion and Future Work}
In this paper, we studied the multiple network access interfaces assisted user cooperation, termed inter-network cooperation, to improve the energy efficiency of cellular uplink transmissions, where a short-range communication network is exploited to assist the cellular transmissions. {By taking into account the upper-layer protocol overhead and physical-layer wireless channel impairments including the path loss, fading, and thermal noise}, we analyzed the energy consumption of traditional scheme without user cooperation, conventional intra-network cooperation, and proposed inter-network cooperation given users' target outage probability and data rate requirements. {It is shown that the energy consumptions of both intra- and inter-network cooperation are reduced with an increasing number of effective bits per data packet.} Numerical results also show that when user terminals are close enough to the base station, the traditional scheme without user cooperation outperforms the intra- and inter-network cooperation in terms of energy consumption. However, as the terminals move away from the base station, the proposed inter-network cooperation substantially reduces energy consumption over the traditional schemes without user cooperation and with intra-network cooperation, which shows the energy saving benefits of inter-network cooperation for cell-edge users when they are not too distant from each other.

{It is worth mentioning that in this paper, we only investigated the inter-network cooperation in a simplified two-user network. It is thus of high practical interest to extend the results of this paper to a general case consisting of more than two users in practical cellular systems. To this end, user pairwise grouping is an efficient solution to addressing the multi-user scenario in which multiple users are divided into multiple pairs of users and different user pairs proceed with the inter-network cooperation independently of each other. It needs to be pointed out that the user pairing in green wireless communications should aim at minimizing the overall energy consumption, differing from existing user grouping strategies in [10] where the focus of user grouping is to improve the transmission reliability (e.g., outage probability minimization). In addition, this paper only examined the impact of upper-layer protocol overhead on the energy consumption of inter-network cooperation without considering the detailed upper-layer protocols and mechanisms. In practical systems, non-negligible energy resources are spent in the upper-layer protocol management, e.g. routing congestion, medium access collision, etc. As a consequence, it is of high importance to explore the inter-network cooperation by jointly considering the network (NET), medium access control (MAC), and physical (PHY) layers in minimizing the overall energy consumption. We will leave the above interesting problems for our future work.}

\ifCLASSOPTIONcaptionsoff
  \newpage
\fi

\appendices
\section{Outage Analysis in The High-SNR Region}
Using (12) and denoting ${\gamma _{1,c}} = {{{P_{1,c}}}}/{{{N_0}{B_c}}}$, ${\gamma _{2,c}} = {{{P_{2,c}}}}/{{{N_0}{B_c}}}$, ${K_1} = {(\dfrac{{{\lambda _c}}}{{4\pi {d_{1b}}}})^2}{G_{U1}}{G_{BS}}$ and ${K_2} = {(\dfrac{{{\lambda _c}}}{{4\pi {d_{2b}}}})^2}{G_{U2}}{G_{BS}}$, we can express $\Pr [\gamma _{1b}^{NC}(\theta  = 1) < ({2^{{{\overline R }_1}/({B_c}\kappa_c)}} - 1)\Delta_c]$ as
\begin{equation}
\begin{split}
&\Pr [\gamma _{1b}^{NC}(\theta  = 1) < ({2^{{{\overline R }_1}/({B_c}\kappa_c)}} - 1)\Delta_c] \\
&= \Pr [{\gamma _{1,c}}{K_1}|{h_{1b}}{|^2} + {\gamma _{2,c}}{K_2}|{h_{2b}}{|^2} < ({2^{{{\overline R }_1}/({B_c}\kappa_c)}} - 1)\Delta_c]\\
 &=\iint\limits_{{{\gamma _{1,c}}{K_1}{x_1} + {\gamma _{2,c}}{K_2}{x_2} < ({2^{{{\overline R }_1}/({B_c}\kappa_c)}} - 1)\Delta_c}}{{\dfrac{1}{{\sigma _{1b}^2\sigma _{2b}^2}}\exp ( - \dfrac{{{x_1}}}{{\sigma _{1b}^2}} - \dfrac{{{x_2}}}{{\sigma _{2b}^2}})d{x_1}d{x_2}}}.
\end{split}\tag{A.1}\label{A.1}
\end{equation}
Denoting ${x_1} = \sigma _{1b}^2{y_1}$ and ${x_2} = \sigma _{2b}^2{y_2}$, (A.1) is rewritten as
\begin{equation}
\Pr [\gamma _{1b}^{NC}(\theta  = 1) < ({2^{{{\overline R }_1}/({B_c}\kappa_c)}} - 1)\Delta_c]= \int_0^{z_1^{ - 1}} {d{y_1}\int_0^{\frac{{1 - {z_1}{y_1}}}{{{z_2}}}} {\exp ( - {y_1} - {y_2})d{y_2}} },
\tag{A.2}\label{A.2}
\end{equation}
where ${z_1} = \frac{{{\gamma _{1,c}}\sigma _{1b}^2{K_1}}}{{({2^{{{\overline R }_1}/({B_c}\kappa_c)}} - 1)\Delta_c}}$ and ${z_2} = \frac{{{\gamma _{2,c}}\sigma _{2b}^2{K_2}}}{{({2^{{{\overline R }_1}/({B_c}\kappa_c)}} - 1)\Delta_c}}$. Letting ${z_1},{z_2} \to \infty $ and using the two-dimensional Taylor expansion series, we have
\begin{equation}
\begin{split}
\exp ( - z_1^{ - 1} - z_2^{ - 1}) = &1 + ( - z_1^{ - 1} - z_2^{ - 1}) + \frac{1}{{2!}}{( - z_1^{ - 1} - z_2^{ - 1})^2} +  \cdots  + \frac{1}{{n!}}{( - z_1^{ - 1} - z_2^{ - 1})^n}\\
&+ O\left( {{{( - z_1^{ - 1} - z_2^{ - 1})}^n}} \right).
\end{split}\tag{A.3}\label{A.3}
\end{equation}
where $O\left(  \cdot  \right)$ represents the higher order infinitesimal. We consider the first two terms in (A.3) and obtain the following approximation
\begin{equation}
\exp ( - z_1^{ - 1} - z_2^{ - 1}) \approx 1 + ( - z_1^{ - 1} - z_2^{ - 1}).\tag{A.4}\label{A.4}
\end{equation}
In order to guarantee a given accuracy $\delta > 0 $ of the Taylor approximation given in (A.4), we obtain
\begin{equation}\nonumber
\frac{1}{{2!}}{( - z_1^{ - 1} - z_2^{ - 1})^2} < \delta ,
\end{equation}
which results in $0 < z_1^{ - 1} + z_2^{ - 1} < \sqrt {2\delta } $ due to ${z_1},{z_2} > 0$. Therefore, considering $0 < z_1^{ - 1} + z_2^{ - 1} < \sqrt {2\delta } $ and using (A.4), we can express (A.2) as
\begin{equation}
\begin{split}
\Pr [\gamma _{1b}^{NC}(\theta  = 1) < ({2^{{{\overline R }_1}/({B_c}\kappa_c)}} - 1)\Delta_c]& = \int_0^{z_1^{ - 1}} {d{y_1}\int_0^{\frac{{1 - {z_1}{y_1}}}{{{z_2}}}} {(1 - {y_1} - {y_2})d{y_2}} } \\
& = \frac{{3 - z_1^{ - 1} - z_2^{ - 1}}}{{6{z_1}{z_2}}}\\
& \approx  \frac{1}{{2{z_1}{z_2}}}.
\end{split}\tag{A.5}\label{A.5}
\end{equation}
where the last approximation arises from the fact that $z_1^{ - 1} + z_2^{ - 1}$ is negligible.

\section{Convex Analysis of (51)}
Let us first review the definitions of concave and convex functions. In general, a continuous and twice-differentiable function is concave, if and only if its Hessian matrix is negative semi-definite. On the contrary, if the Hessian matrix is positive semi-definite, the function is convex. Then, the following discuses the convexity of the two inequality constraints $2{\overline {{\rm{Pout}}} _1}{z_1}{z_2} - 2(1 - \alpha ){z_2} - \alpha  \ge 0$ and $2{\overline {{\rm{Pout}}} _2}{\beta ^2}{z_1}{z_2} - 2(1 - \alpha ){\beta ^2}{z_1} - \alpha \ge 0 $ in (51). Denoting $f({z_1},{z_2}) = 2{\overline {{\rm{Pout}}} _1}{z_1}{z_2} - 2(1 - \alpha ){z_2} - \alpha $, we can compute the Hessian matrix of $f({z_1},{z_2})$ as
\begin{equation}
{H_f} = \left[ {\begin{array}{*{20}{c}}
{\dfrac{{{\partial ^2}f({z_1},{z_2})}}{{\partial z_1^2}}}&{\dfrac{{{\partial ^2}f({z_1},{z_2})}}{{\partial {z_1}\partial {z_2}}}}\\
{\dfrac{{{\partial ^2}f({z_1},{z_2})}}{{\partial {z_1}\partial {z_2}}}}&{\dfrac{{{\partial ^2}f({z_1},{z_2})}}{{\partial z_2^2}}}
\end{array}} \right] = \left[ {\begin{array}{*{20}{c}}
0&{2{{\overline {{\rm{Pout}}} }_1}}\\
{2{{\overline {{\rm{Pout}}} }_1}}&0
\end{array}} \right]. \tag{B.1}\label{B.1}
\end{equation}
Thus, the determinants of all leading principal minors of the above matrix are given by
\begin{equation}
\begin{split}
&|H_f^1| = \left| 0 \right| = 0\\
&|{H_f^2}| = \left| {\begin{array}{*{20}{c}}
0&{2{{\overline {{\rm{Pout}}} }_1}}\\
{2{{\overline {{\rm{Pout}}} }_1}}&0
\end{array}} \right| =  - 4{({\overline {{\rm{Pout}}} _1})^2} < 0,
\end{split}\tag{B.2}\label{B.2}
\end{equation}
which shows that ${H_f}$ is an indefinite matrix, implying that $f({z_1},{z_2}) = 2{\overline {{\rm{Pout}}} _1}{z_1}{z_2} - 2(1 - \alpha ){z_2} - \alpha $ is neither concave nor convex and $2{\overline {{\rm{Pout}}} _1}{z_1}{z_2} - 2(1 - \alpha ){z_2} - \alpha  \ge 0$ is thus a non-convex constraint. Similarly, denoting $g({z_1},{z_2}) = 2{\overline {{\rm{Pout}}} _2}{\beta ^2}{z_1}{z_2} - 2(1 - \alpha ){\beta ^2}{z_1} - \alpha $, we can obtain the Hessian matrix of $g({z_1},{z_2})$ as
\begin{equation}
{H_g} = \left[ {\begin{array}{*{20}{c}}
0&{2{{\overline {{\rm{Pout}}} }_2}{\beta ^2}}\\
{2{{\overline {{\rm{Pout}}} }_2}{\beta ^2}}&0
\end{array}} \right], \tag{B.3}\label{B.3}
\end{equation}
from which the determinants of leading principal minors of $H_g$ are obtained as
\begin{equation}
|H_g^1| = 0 {\textrm{  and  }}|{H_g^2}| =  - 4{({\overline {{\rm{Pout}}} _2})^2}{\beta ^4} < 0, \tag{B.4}\label{B.4}
\end{equation}
which implies that $2{\overline {{\rm{Pout}}} _2}{\beta ^2}{z_1}{z_2} - 2(1 - \alpha ){\beta ^2}{z_1} - \alpha  \ge 0$ is non-convex. Since both the inequality constraints $2{\overline {{\rm{Pout}}} _1}{z_1}{z_2} - 2(1 - \alpha ){z_2} - \alpha  \ge 0$ and $2{\overline {{\rm{Pout}}} _2}{\beta ^2}{z_1}{z_2} - 2(1 - \alpha ){\beta ^2}{z_1} - \alpha  \ge 0$ are non-convex, one can easily conclude that the formulated optimization problem in (51) is non-convex.

\end{document}